\documentclass[11pt]{article}
\usepackage{amsmath,amsthm,amssymb}
\usepackage[margin=.93in]{geometry}
\usepackage[nodisplayskipstretch]{setspace}
\usepackage[hypertexnames=false]{hyperref}
\hypersetup{
    colorlinks,
    linkcolor={blue!30!black},
    citecolor={blue!30!black},
    urlcolor={blue!30!black},
    breaklinks=true
}
\usepackage{xurl}
\usepackage{cleveref}
\usepackage{xcolor}
\usepackage{graphicx}
\usepackage{caption}
\usepackage{placeins}
\usepackage{xspace}
\usepackage{csquotes}
\usepackage[footfont=normalsize,font=normalsize]{floatrow}
\usepackage{ragged2e}
\usepackage{tabularx}
\usepackage{booktabs}
\usepackage{mdframed}
\usepackage{xr}
\usepackage{relsize}
\usepackage[T1]{fontenc}
\usepackage{lmodern}
\usepackage[bottom]{footmisc}
\usepackage[bibstyle=numeric, citestyle=authoryear-comp, doi=false, url=false, backend=biber, maxbibnames=6, maxcitenames=4, uniquelist=false, uniquename=false, sorting=nyt, natbib=true, style = authoryear-comp]{biblatex}
\renewbibmacro{in:}{}
\usepackage{tikz}
\usetikzlibrary{shapes,positioning,intersections,quotes,arrows.meta,bending}

\newcommand{\point}[1]{}

\newtheorem{theorem}{Theorem}
\newtheorem{assumption}{Assumption}

\newtheorem{corollary}{Corollary}
\newtheorem{remark}{Remark}
\newtheorem{inneruassumption}{Assumption}
\newtheorem{algorithm}{Algorithm}
\newenvironment{namedassumption}[1]
  {\renewcommand\theinneruassumption{#1}\inneruassumption}
  {\endinneruassumption}

\crefname{namedtheorem}{Theorem}{Theorems}

\crefname{assumption}{Assumption}{Assumptions}
\crefname{namedassumption}{Assumption}{Assumptions}
\crefname{lemma}{Lemma}{Lemmas}
\newcommand{\E}{\mathbb{E}}

\newcommand{\Var}{\mathrm{Var}}
\newcommand{\indicator}[1]{\mathbf{1}\{#1\}}
\renewcommand{\P}{\mathbb{P}}
\newcommand{\T}{T}

\newcommand{\ATT}{ATT} %{ATT_{D>0}}
\newcommand{\ACRT}{ACRT} %{ACRT_{D>0}}
\newcommand{\LATT}{ATT}
\newcommand{\LACRT}{ACRT}
\newcommand{\loc}{\texttt{loc}}
\newcommand{\glob}{\texttt{glob}}

\allowdisplaybreaks
\newcommand{\alphahat}{\widehat{\alpha}}
\newcommand{\Khat}{\widehat{{K}}}

\AtBeginDocument{
\addtolength{\abovedisplayskip}{-1ex}
\addtolength{\belowdisplayskip}{-1ex}
\setlength{\abovedisplayskip}{0cm}
\setlength{\belowdisplayskip}{0pt}
\setlength{\abovedisplayshortskip}{0cm}
\setlength{\belowdisplayshortskip}{0cm}
}

\usepackage{etoolbox}
\newcommand{\zerodisplayskips}{%
  \setlength{\abovedisplayskip}{7pt}%
  \setlength{\belowdisplayskip}{6pt}%
  \setlength{\abovedisplayshortskip}{5pt}%
  \setlength{\belowdisplayshortskip}{5pt}}
\appto{\normalsize}{\zerodisplayskips}
\appto{\small}{\zerodisplayskips}
\appto{\footnotesize}{\zerodisplayskips}

\numberwithin{equation}{section}
\numberwithin{theorem}{section}
\numberwithin{proposition}{section}
\numberwithin{corollary}{section}
\numberwithin{lemma}{section}
\numberwithin{remark}{section}

\title{Difference-in-Differences with a Continuous Treatment\thanks{We thank the participants of many seminars, workshops, and conferences for their comments. We are grateful to Xiaohong Chen for numerous discussions about implementing the data-driven sieve estimator used in this paper, Amy Finkelstein for sharing her data with us, Carol Caetano, Greg Caetano, Stefan Hoderlein, Jo Mullins, Jon Roth, and Abbie Wozniak for their comments, and Honey Batra for valuable research assistance. Code implementing the methods proposed in the paper is available in the R package \texttt{contdid}, which is available on CRAN.  The views expressed here are those of the authors and do not necessarily represent those of the Federal Reserve Bank of Minneapolis or the Federal Reserve System. The Supplementary Appendix is available \href{https://psantanna.com/files/CGBS_supp_v4.pdf}{here}.}}
\author{Brantly Callaway\thanks{University of Georgia.  Email: \href{mailto:brantly.callaway@uga.edu}{brantly.callaway@uga.edu}} \and Andrew Goodman-Bacon\thanks{Federal Reserve Bank of Minneapolis and NBER. Email: \href{mailto:andrew@goodman-bacon.com}{andrew@goodman-bacon.com}} \and Pedro H. C. Sant'Anna\thanks{Emory University.  Email: \href{mailto:pedro.santanna@emory.edu}{pedro.santanna@emory.edu}}}

\date{First draft on arXiv: July 6, 2021. This draft: December 31, 2025}

\begin{document}

\begin{titlepage}
\clearpage
\maketitle
\thispagestyle{empty}

\begin{abstract}
    \noindent This paper analyzes difference-in-differences designs with a continuous treatment. We show that treatment-on-the-treated-type parameters are identified under a parallel trends assumption analogous to the binary treatment case. However, comparing these parameters across treatments is challenging because parallel trends does not rule out selection bias. We discuss alternative, typically stronger, assumptions that eliminate selection bias. We further show that popular two-way fixed effects estimands admit multiple interpretations, depending on the underlying causal building block, all having important limitations as meaningful summaries of treatment effects. Finally, we introduce alternative estimation procedures that avoid these drawbacks and demonstrate them in an empirical application.
\end{abstract}

\vspace{50pt}


\bigskip


\vspace{100pt}

\end{titlepage}
\pagebreak

\onehalfspacing

\section{Introduction}
The canonical difference-in-differences (DiD) research design compares outcomes before and after treatment started (difference one), between treated and untreated groups (difference two). However, in many DiD applications, the treatment does not simply ``turn on'', it has a ``dose'' or operates with varying intensity. 
Pollution dissipates across space, affecting locations near its source more severely than faraway locations. Localities spend different amounts on public goods and services, or set different minimum wages. Students choose how long to stay in school. 

Continuous treatments can offer advantages over binary ones.\footnote{We generally use ``continuous'' treatments also to mean multi-valued ordered discrete treatments, but make the distinction explicit for certain results.} Variation in intensity makes it possible to evaluate treatments that all units receive.  A clear ``dose-response'' relationship between outcomes and treatment intensity can bolster the case for a causal interpretation or test a theoretical prediction. Finally, we may care more about the effect of changes in treatment intensity, such as increased funding, pollution abatement, or expanded eligibility, than about the effect of the existence of a treatment that already exists.

Despite how conceptually useful and practically common continuous DiD designs are, currently available econometric results provide little guidance on applying and interpreting them, except in some specific cases.  In this paper, we introduce a set of tools that are suitable for DiD setups with variation in treatment dosage. In particular, we (a) discuss how one can identify a variety of treatment effect parameters by exploiting parallel-trends-type assumptions, (b) demonstrate that a simple linear two-way fixed effects (TWFE) estimand accommodates \textit{multiple} decompositions that are difficult to justify as meaningful summaries of treatment effects, and (c) propose ``forward-engineered'' estimators that directly target well-defined causal objects, allowing for more transparent interpretation and robust inference in applications with treatment effect heterogeneity.

To foster intuition and simplify exposition, we start by discussing causal parameters in a two-period DiD design in which units move from no treatment to a non-zero dose. We call the difference between a unit's potential outcome under dose $d$ and its untreated potential outcome a \emph{level treatment effect}. We call the difference in a unit's potential outcome with a marginal increase in the dose a \emph{causal response} \citep{angrist-imbens-1995}. 
When treatment is binary, these two notions of treatment effects coincide, but they do not under a continuous treatment. Importantly, level treatment effects and causal responses can have meaningfully different interpretations, and we establish that they generally require different identifying assumptions as well. 

Comparisons between treated and untreated units identify average (level) treatment effect parameters under a parallel trends assumption on untreated potential outcomes, similar to binary DiD designs. Comparisons between adjacent dose groups, however, do not identify average causal response parameters under the ``standard'' parallel trends assumption. They include causal responses but are contaminated by an additional term involving possibly different treatment effects of the same dose for different dose groups---we refer to this additional term as selection bias.\footnote{In applications where units choose their amount of the treatment, it is natural to refer to this term as selection bias.  In other applications where the dose measures a unit's amount of exposure to some treatment, a different term, such as ``heterogeneity bias'', could be more appropriate.  For simplicity, throughout the paper, we simply refer to this term as selection bias.} We discuss an alternative, typically stronger assumption, which we call strong parallel trends, that says that the average evolution of outcomes for the entire treated population under dose $d$ is equal to the path of outcomes that dose group $d$ actually experienced. Thus, strong parallel trends justifies comparing dose groups by restricting treatment effect heterogeneity. Strong parallel trends may not be plausible in many applications.  Currently, in empirical work, it is common for papers to write as if they have assumed standard parallel trends and interpret their results as if they have assumed strong parallel trends.  Our results clarify what causal questions can be answered under standard parallel trends and what causal questions require stronger assumptions. 

The ideas discussed above are in the spirit of what \citet{mogstad-torgovitsky-2024} call \textit{forward engineering}, where the researcher clearly specifies target parameters and assumptions up front and builds estimators to implement the identification strategy.  Our second main contribution is to \textit{reverse engineer}  the most common way that practitioners estimate continuous DiD designs, which is to run a TWFE regression that includes time fixed effects ($\theta_t$), unit fixed effects ($\eta_i$), and the interaction of a dummy for the post-treatment period ($Post_t$) with a variable that measures unit $i$'s dose or treatment intensity, $D_{i}$:
\vspace{-.1cm}\begin{align}\label{eqn:twfe}
    Y_{i,t} = \theta_t + \eta_i + \beta^{twfe} D_i\cdot Post_t + v_{i,t}. 
\end{align}
This TWFE specification is clearly motivated by DiD setups with two periods and two treatment groups, though many prominent textbooks suggest using it in more general setups (e.g., \citealp{cameron-trivedi-2005}, \citealp{angrist-pischke-2008}, and \citealp{wooldridge-2010}). There are several ways to interpret $\beta^{twfe}$, each corresponding to a different type of causal parameter. We decompose it in terms of level effects, scaled level effects, causal responses, and scaled high-versus-low ($2\times 2$) effects.  Each decomposition is a weighted integral of dose-specific causal parameters, and none provides a clear causal and policy-relevant interpretation of $\beta^{twfe}$, at least not when treatment effects are allowed to vary across doses and/or groups.\footnote{The decompositions that we provide are specific to the particular TWFE regression specification in \Cref{eqn:twfe}, which we focus on due to its ubiquity in empirical work with a continuous treatment. Some of the drawbacks we discuss below, particularly regarding weighting schemes inherited from the TWFE regression, could be addressed by considering a more flexible specification. See, e.g., \citet{wooldridge-2021} for a discussion involving binary treatments.\label{fn:flexible_twfe}}

For instance, we show that $\beta^{twfe}$ can be expressed as a weighted integral of average level treatment effect parameters but where the weights integrate to zero, indicating that $\beta^{twfe}$ should not be interpreted as an average (level) treatment effect.  Interestingly, however, TWFE puts negative weights on the below-average dose units and positive weights on above-average dose units, and, thus, after re-scaling by a weighted average of the difference between doses for high- and low-dose units, is equivalent to a weighted binary DiD using higher-dose units as the ``treated'' group and lower-dose units as the ``comparison'' group, with weights proportional to a unit's absolute distance from the mean dose. Our next decomposition, based on average level treatment effect parameters scaled by their dose, also displays negative weights, though their weights integrate up to one and not zero. 

In contrast, a TWFE decomposition in terms of average causal response parameters has weights that integrate up to one and are non-negative, but also includes a selection bias term stemming from effect heterogeneity across doses. The strong parallel trends assumption eliminates this selection bias. The weights on causal responses at different doses, however, differ from the distribution of the dose among the treated, which creates a further challenge to interpreting $\beta^{twfe}$, even if strong parallel trends holds. The weights are also undesirably sensitive to the size of the untreated group.  In our application, if we drop the untreated group, which changes the weights but does not change the underlying average causal responses, our estimate of $\beta^{twfe}$ shrinks by 78\%. Our decomposition of $\beta^{twfe}$ based on scaled $2\times 2$ average effects as building blocks also highlights limitations of using $\beta^{twfe}$ as a causal summary parameter.

We demonstrate that these drawbacks are easily avoidable and discuss different DiD estimators that build upon our identification results and recover interpretable causal parameters. When the treatment is discrete, this is as simple as running a linear regression with multiple treatment indicators, which is similar to staggered DiD setups \citep{callaway-santanna-2021}. When the treatment is continuous, there are several options, including adopting a parametric, semiparametric, or nonparametric regression model. In particular, we discuss how to adapt the data-driven sieve-based nonparametric regression proposed by \citet{chen-christensen-kankanala-2024} to our context, although we note that other semi/nonparametric procedures are also possible. We also show how to construct causal summary measures of our average treatment effect functions that bypass the TWFE weighting problems by using the dose density as weights. Our results suggest that one can easily summarize average level treatment effects among treated units by comparing the average change in outcomes for all treated units to the average change in outcomes for untreated units. This can be estimated by running a binary DiD with a ``treatment dummy'' equal to one for any units with positive doses. Summarizing average causal responses using dose density weights involves estimating an average derivative, which is simple to compute using ``flexible'' linear regressions. We also discuss how to construct event-study results using these summary measures, which can then be used to assess the plausibility of the parallel trends assumptions.

To contrast our proposed estimators with TWFE in practical settings, we revisit \citeauthor{acemoglu-finkelstein-2008}'s \citeyear{acemoglu-finkelstein-2008} study of a 1983 Medicare reform that eliminated labor subsidies for hospitals. The original paper uses a TWFE estimator to compare the change in capital-labor ratios between hospitals whose input prices were more or less affected by the end of the subsidy. It concludes that price regulations favoring capital significantly increase capital use. The distinction between level treatment effect parameters and causal responses is important in this example: a positive level treatment effect shows that the policy as a whole increased the use of capital, while causal responses describe which subsidy levels generated the largest responses. We find that the reform raised capital-labor ratios by about 18 percent (on average), which is 50 percent larger than the comparable TWFE estimate because of the weighting issues highlighted by our decompositions. We also estimate variable average causal response ($\ACRT$) parameters that are quite large at low subsidy levels---implying elasticities of substitution greater than 2---yet slightly \textit{negative} for most positive doses. These negative $\ACRT$ estimates cast doubt about the plausibility of the strong parallel trends assumption, the simple two-factor model of hospital production, or both. Our results support \citeauthor{acemoglu-finkelstein-2008}'s \citeyear{acemoglu-finkelstein-2008} conclusion that the 1983 Medicare reform led hospitals to favor capital over labor, but suggest caution in a policy interpretation about which subsidy levels have the largest effects or an economic interpretation in terms of production function parameters. 
\medskip

\noindent \textbf{Related Literature:} Our paper contributes to the literature on modern DiD methods; see, e.g., \citet{baker-callaway-cunningham-goodman-santanna-2025} for an overview. We contribute to this literature by highlighting challenges associated with using TWFE with continuous treatments, discussing the role of different parallel trends assumptions to learn about different causal parameters, and providing easy-to-use estimation procedures that can highlight treatment effect heterogeneity with continuous treatments. 

The closest paper to ours is \citet{fricke-2017}, which focuses on DiD setups with two time periods and three treatment dosages: H, L, and 0. \citet{fricke-2017} shows that under standard parallel trends, one can identify dose-specific average treatment effect among dose groups in two-period DiD designs. He also considers stronger assumptions that permit causal interpretation of the $H$ vs.~$L$ contrast when an untreated group is not available. We generalize his identification results to DiD settings with richer treatment distributions, including continuous cases, multiple time periods, and staggered treatment adoptions. This allows us to (i) discuss a broader set of parameters of interest that are suitable for incremental changes in treatment dose, (ii) discuss event-study and other types of treatment aggregations,  (iii) derive decomposition results that question the causal meaning of TWFE estimates under treatment effect heterogeneity, and (iv) offer identification-based estimation templates for researchers to avoid the pitfalls of simple TWFE specifications with continuous/multi-valued treatments.

Our paper is also related to %\citet{chaisemartin-dhaultfoeuille-2018}, \citet{chaisemartin-dhaultfoeuille-2024} and \citet{chaisemartin-dhaultfoeuille-pasquier-sow-vazquez-2025}
a series of papers on more complicated non-binary DiD setups. \citet{chaisemartin-dhaultfoeuille-2018} focuses on fuzzy designs, where a researcher is interested in individual-level effects of a binary treatment that has been aggregated across units into a continuous ``treatment rate.'' In contrast, we study ``sharp'' designs in which the treatment exposure is itself continuous or multi-valued discrete at the unit-level. The approach proposed in \citet{chaisemartin-dhaultfoeuille-2024} can also accommodate continuous treatments, although they focus on aggregated target parameters rather than dose-specific estimators of treatment effect heterogeneity, as we do. They also do not discuss identification, nor estimation, related to average causal response parameters, an important focus of our analysis. \citet{chaisemartin-dhaultfoeuille-pasquier-sow-vazquez-2025} builds on \citet{chaisemartin-dhaultfoeuille-2024} and considers a DiD setup with continuous treatments with potentially non-staggered (but static) treatments. Our proposal differs in its target parameters and DiD designs. %While we consider both  functional parameters (dose-response and average causal response curves) and causal summary measures, \citet{chaisemartin-dhaultfoeuille-pasquier-sow-vazquez-2025} focus on distance-weighted) averages of what we refer to as $2\times2$ average effects. 
Similarly to \citet{chaisemartin-dhaultfoeuille-2024}, \citet{chaisemartin-dhaultfoeuille-pasquier-sow-vazquez-2025} average effects of discrete rather than marginal changes of treatments. On the other hand, \citet{chaisemartin-dhaultfoeuille-pasquier-sow-vazquez-2025} allows for units to already be exposed to the treatment in the first period and considers instrumental variable extensions, which we do not. 

Our decompositions also relate to the literature on TWFE bias in heterogeneous treatment effect settings with a binary treatment (e.g., \citealp{goodman-bacon-2021,chaisemartin-dhaultfoeuille-2020, sun-abraham-2021, borusyak-jaravel-spiess-2024}) but we extend this logic to continuous treatments and highlight that the same TWFE regression coefficient can have multiple interpretations depending on the ``building blocks'', and that new ``bias'' terms may appear, depending on the type of parallel trends assumption being used. A perhaps more practically relevant message from our decompositions is that, even when all weights are non-negative, TWFE can still provide an unappealing causal summary parameter. Interestingly, if one replaces our DiD setting with a cross-sectional design with a randomly assigned dose, all four of our decomposition results would still hold, highlighting that linear specifications may not be desirable with continuous treatments, even when the dose is fully randomized. 

Finally, we note that some of our causal response decomposition builds on \citet[Proposition 2]{yitzhaki-1996}, which expresses the slope coefficient in a regression of an outcome on a continuous variable as a weighted average of underlying local slopes.  Besides differences related to causal interpretations and panel data, we mildly extend those results to accommodate a mass of untreated units.

\section{Motivating Continuous DiD from an Empirical Perspective} \label{sec:background}
To fix ideas and provide intuition for our results, we revisit \citeauthor{acemoglu-finkelstein-2008}'s \citeyear{acemoglu-finkelstein-2008} (AF) study of how price regulations affect firms' input choices. When Medicare began in 1965, hospitals received reimbursements from the federal government for a share of their labor and capital expenditures proportional to the fraction of total patient days accounted for by Medicare recipients ($m_i$). Hospital $i$ thus faced input prices equal to $(1-s_Lm_i)w$ for labor and $(1-s_Km_i)r$ for capital, where $s_L$ and $s_K$ are the labor and capital subsidy rates and $w$ and $r$ are market wages and rental rates. In 1983, Medicare moved to the Prospective Payment System (PPS), which replaced the labor subsidy with a small payment per episode/diagnosis. This set $s_L=0$ but left the capital subsidy unchanged. Therefore, the price of labor for a given hospital rose from $(1-s_Lm_i)w$ to $w$, skewing relative factor prices.

The statutory relationship between a hospital's Medicare volume, $m_i$, and the change in its price of labor, $s_Lm_iw$, motivates AF's use of a continuous DiD design comparing changes in capital/labor ratios before and after 1983 between hospitals with different pre-PPS Medicare inpatient shares. AF's description, estimation, and interpretation of this empirical strategy touch on some of the most common ways of justifying and implementing continuous DiD designs.

One motivation for this design is practical: variation in a dose (or exposure) permits the evaluation of treatments for which binary DiD is either infeasible or undesirable. In AF's case, about 15 percent of hospitals were ``untreated'' by the change in Medicare's subsidy policy because they served non-Medicare-eligible populations, like children or psychiatric patients, so one may be concerned about whether these constitute a valid comparison group. AF therefore describe $m_i$, which is the hospital's Medicare volume in 1983, as an ``attractive source of variation'' in the price of labor both because it varies substantially---the mean of $m_i$ among treated hospitals is 0.45, and the standard deviation is 0.15---and because hospitals with $m_i>0$ may be more comparable to each other than treated hospitals are to untreated hospitals.

Another common justification for continuous DiD designs is that a ``dose-response'' relationship between exposure and outcomes can support a causal interpretation or test a theoretical prediction. \citet[p.\,158]{meyer-1995}, for example, argues that ``differences in the intensity of the treatment across different groups allow one to examine if the changes in outcomes differ across treatment levels in the expected direction.'' AF lay out a simple theoretical framework in which the move to PPS should (i) raise capital/labor ratios and (ii) do so more strongly for hospitals with higher pre-PPS values of $m_i$. They view their continuous DiD design as a way to estimate a causal effect of PPS as a whole and test the theoretical predictions of their model.

Finally, researchers often advocate for continuous DiD designs because they can be used to estimate average causal effects of small changes in the dose. In many economic models, price and income elasticities determine optimal policies like tax rates, tax bases, subsidies, and regulations \citep{hendren-2016}, but these are continuous concepts that can be estimated accurately only with continuous variation. AF's theoretical framework implies that, under some assumptions, DiD estimates provide information about hospitals' elasticity of substitution between capital and labor, although AF do not argue for this kind of ``marginal'' interpretation.

\begin{figure}[ht]
   \begin{center}
     \caption{Two-Way Fixed Effects Event-Study Estimates of the Effect of Medicare's Reimbursement Reform on Hospital Input Mix}
    \includegraphics[width=.75\textwidth, keepaspectratio]{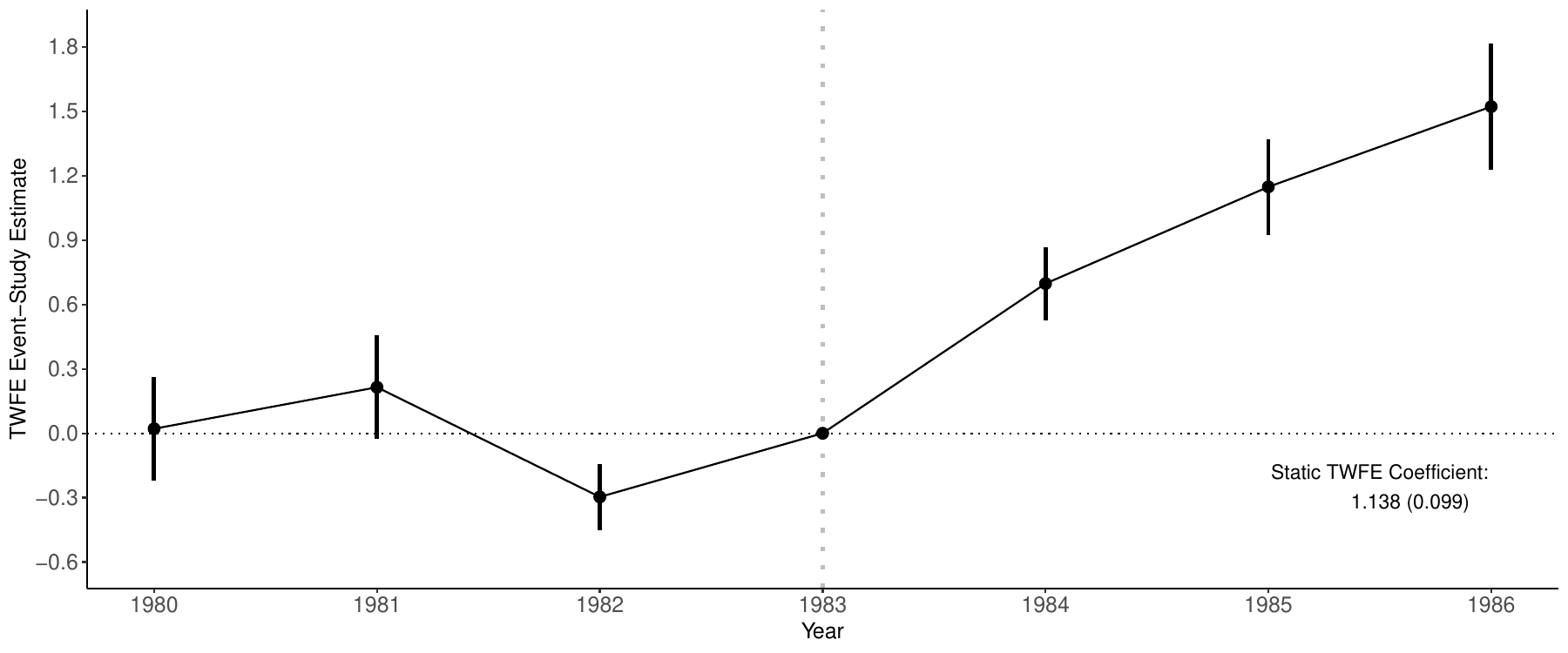}
   \label{fig:F1}
   \end{center}
   \justifying
   \setstretch{.75}
   {\vspace{-.5cm} \scriptsize  \textit{Notes:} The figure plots TWFE event-study coefficients and their 95\% confidence intervals from regressions with hospital fixed effects, year fixed effects, and the 1983 Medicare inpatient share ($m_i$) interacted with either a dummy for years after 1983 (static TWFE specification) or the year dummies (event study). The outcome variable is the depreciation share of total operating expenses, a measure of hospitals' capital/labor ratio. The data cover the years 1980-1986 and come from the American Hospital Association's annual survey \citep{aha-1986}. The results are not numerically identical to AF's because we drop 860 hospitals (out of 6,741) with missing outcomes for some years.}% We dropped 860 hospitals (out of 6741) that have missing data for the outcome. We also report the static TWFE coefficient and standard errors associated with \eqref{eqn:twfe}. All standard errors are clustered at the hospital level.}
\end{figure}

In terms of estimation, AF use the standard tool for continuous DiD designs: a TWFE regression with hospital and year fixed effects. They follow textbook advice. \citet[p.\,132]{wooldridge-2010} observes that a two-period DiD regression estimator ``can be easily modified to allow for continuous, or at least non-binary, `treatments'.''  \citet[p.\,234]{angrist-pischke-2008} emphasize ``a second advantage of regression DD is that it facilitates the study of policies other than those that can be described by a dummy.'' They also follow common practice and describe their identifying assumption as an extension of the parallel trends assumption from binary designs: ``\emph{Without the introduction of PPS}, hospitals with different $m_i$'s would not have experienced differential changes in their outcomes in the post-PPS period'' (emphasis added).

\Cref{fig:F1} reproduces AF's DiD event-study coefficients for each calendar year, relative to 1983, and the estimate of $\beta^{twfe}$ from an equation like \eqref{eqn:twfe}. AF interpret these results as indicative that after 1983, capital/labor ratios rose more strongly for hospitals with higher values of $m_i$, without a substantial differential change in input mix before PPS. Our impression is that event-study results like those in \Cref{fig:F1} would usually be interpreted as strong causal evidence because there are (relatively) small pre-trend estimates, large estimates in post-treatment periods, %large differences in outcomes between higher- and lower-dose units after treatment, 
and tight confidence intervals. What is missing from most continuous DiD analyses, however, is a specific statement about \emph{what} causal parameters researchers would like to estimate, the assumptions under which they are identified, and a formal justification for a particular estimator. Our goal is to shed light on these three issues. 

\section{Baseline Case: A New Treatment with Two Periods} \label{sec:baseline}
We illustrate our main points in a setup with two periods of panel data, $t=1$ and $t=2$. In the first period, no unit is treated. In the second period, some units receive a treatment ``dose,'' denoted by $D_i$, and others remain untreated. Extensions to multiple periods and staggered setups are discussed in \Cref{sec:extensions}.  We denote the support of $D$ by $\mathcal{D}$. $D_i$ can be (absolutely) continuous or can be multi-valued discrete, but to simplify the exposition, we refer to it as ``continuous.'' We define potential outcomes for unit $i$ in period $t$ as $Y_{i,t}(0,d)$, where potential outcomes are indexed by the treatment sequence \citep{robins_new_1986}. As we focus on the setup where all units have $d=0$ in period $t=1$, we simplify the potential outcome notation and henceforth write $Y_{i,t}(d)$, where $d$ is the dosage in period $t=2$. This is the outcome that unit $i$ would experience in period $t$ under (period-two) dose $d$. In each time period $t$, the observed outcome for unit $i$ is $Y_{i,t}=Y_{i,t}(D_i)$. We assume that all expectations are finite and well-defined. Henceforth, we omit the unit index $i$ to make the notation less cluttered and define $\Delta Y = Y_{t=2}-Y_{t=1}$.

\subsection{Parameters of Interest with a Continuous Treatment}

The potential outcomes notation $Y_t(d)$ reflects that treatment can take many values, and so each unit can experience many types of causal effects. The \emph{level treatment effect} of dose $d$ in time period $t$ for a given unit is defined as its potential outcome when $D=d$ minus its untreated potential outcome: $Y_{t}(d) - Y_{t}(0)$. Level treatment effects measure the treatment effect at time $t$ from switching treatment dosage from $0$ to $d$. This is a straightforward extension of a binary ``treatment effect'' to a continuous ``treatment effect function'' or ``dose-response function.'' 

But zero-treatment is not the only relevant counterfactual. We define a unit's \emph{causal response} at $d$ as $Y_t'(d)$, the derivative of the potential outcome with respect to dose $d$ (when the treatment is continuous),\footnote{This is a slight abuse of notation as we do not require $Y_{t}(d)$ to be differentiable (or even continuous), but rather we mean here the causal effect of a marginal increase in the dose on a unit's outcome: $\lim_{h \rightarrow 0^+} (Y_{t}(d+h) - Y_{t}(d))/h$.} or as the difference in potential outcomes between adjacent doses scaled by the difference in the doses, $\big(Y_t(d_j) - Y_t(d_{j-1})\big)\big/(d_j - d_{j-1})$ (when the treatment is discrete). Causal responses measure the treatment effect at time $t$ of a marginal increment of dose $d$. These two types of treatment effects---the level of $Y_{t}(d) - Y_{t}(0)$ or its slope, $Y^\prime_{t}(d)$---define unit-level causal parameters in continuous designs, and connect to results in the instrumental variables (IV) literature on multi-valued discrete or continuous endogenous variables (\cite{angrist-imbens-1995}, \cite{angrist-graddy-imbens-2000}).

We focus on ``building block'' parameters that are averages of these two kinds of causal effects in the post-treatment period, $t=2$. Average level treatment effects (which we refer to as average treatment effects) extend definitions from the binary case:
\begin{align*}
    \LATT(d|d^\prime) = \E[Y_{t=2}(d) - Y_{t=2}(0) | D=d^\prime] \quad \textrm{and} \quad \ATT(d) = \E[Y_{t=2}(d) - Y_{t=2}(0) | D>0],
\end{align*}
where $\LATT(d|d^\prime)$ is the average effect of dose $d$ compared to zero dosage in the post treatment period $t=2$ on units that actually experienced dose $d^\prime$. When $d^\prime=d$, this is the $\LATT$ local to units that received dose $d$.  $\ATT(d)$ is the average difference between potential outcomes under dose $d$ relative to untreated potential outcomes across all treated units, not just those that experienced dose $d$, in time period $t=2$.

Average causal response parameters for absolutely continuous treatments are defined as
\begin{small}
\begin{align*}
    \LACRT(d|d^\prime) = \frac{\partial \LATT(l|d^\prime)}{\partial l} \bigg|_{l=d} =  \frac{\partial \E[Y_{t=2}(l)| D=d^\prime]}{\partial l} \bigg|_{l=d} \textrm{ and } \ACRT(d) =\frac{\partial \ATT(d)}{\partial d}= \frac{\partial \E[Y_{t=2}(d)|D>0]}{\partial d}.
\end{align*}
\end{small}
$\LACRT(d|d')$ is the average effect of a marginal dose increase from $d$ for dose group $d'$.  It equals the derivative of $\LATT(l|d')$ with respect to $l$, evaluated at $l=d$, which is equivalent to the derivative of the $t=2$ average potential outcome function with respect to dose $d$ among dose group $d'$. $\ACRT(d)$ is the average causal response of dose $d$ across all treated units.  For discrete treatments, average causal responses are defined in a similar way but with slightly different notation:
\begin{small}
\begin{align*}
    \LACRT(d_j|d_k) = \frac{\E[Y_{t=2}(d_j) - Y_{t=2}(d_{j-1}) | D=d_k]}{d_j-d_{j-1}} \quad \textrm{and} \quad \ACRT(d_j) = \frac{\E[Y_{t=2}(d_j) - Y_{t=2}(d_{j-1})|D>0]}{d_j - d_{j-1}} .
\end{align*}
\end{small}
$\LACRT(d_j|d_k)$ equals the difference in mean potential outcomes between dose level $d_j$ and the next lowest dose $d_{j-1}$ in period $t=2$ for dose group $d_k$, scaled by the difference between the two doses.  Similarly, $\ACRT(d_j)$ gives the average causal response of dose $d_j$ relative to dose $d_{j-1}$, but it is for the entire treated group. We note that our definition of $ \LACRT(d_j|d_k)$ and $\ACRT(d_j)$ differs from the definitions in \citet{angrist-imbens-1995}, as it scales the changes in expected potential outcomes by the change in dosage.  %We follow the literature, particularly \citet{angrist-imbens-1995}, by not defining $\LACRT(d_j|d_j)$ as being scaled by the difference between $d_j$ and $d_{j-1}$ though, up to definitions of parameters, that does not affect the results below.

\begin{figure}[!ht]
   \begin{center}
   \caption{{Causal Parameters in a Continuous Difference-in-Differences Design}}
\begin{tikzpicture}[scale=.7]
%x-axis
  \draw[->] (0, 0) -- (10, 0) node[right] {\scriptsize $l$};
%y-axis (take out label)
  \draw[->] (0, 0) -- (0, 8);
  %{$Y_{t}(d)-Y_{t}(0)$}
%curve--kind of cool how you define the domain, the variable name, and the function!
 \draw[scale=1.9, domain=0:4.8, smooth, variable=\d, black,line width=0.4mm] plot ({\d}, {ln(1+8*\d)-.1*(\d-4.5)*(\d>4.5)});
%curve label 
 \node[circle,inner sep=0pt,fill=white,label=below right:{\tiny $\LATT(l|d)$}] at (4.2*1.9,3.418*1.9) {};

%labels for points on the y-axis
 \node[circle,inner sep=0pt,fill=white,label=left:{\tiny $\LATT(d'|d)$}] at (0,3.497*1.9) {};
 \node[circle,inner sep=0pt,fill=white,label=left:{\tiny $\LATT(d|d)$}] at (0,2.197*1.9) {};
%labels for points on the x-axis
 \node[circle,inner sep=0pt,fill=black,label=below:{\tiny $d'=d+1$}] at (4*1.9,0) {};
 \node[circle,inner sep=0pt,fill=black,label=below:{\tiny $d$}] at (1*1.9,0) {}; 
%vertical line at d
 \draw[densely dotted] (1*1.9,0) -- (1*1.9,2.197*1.9);
%vertical line at d' 
 \draw[densely dotted] (4*1.9,0) -- (4*1.9,3.497*1.9);
%horizontal line at \LATT(d'|d) 
 \draw[densely dotted] (0,3.497*1.9) -- (4*1.9,3.497*1.9);
%horizontal line at \LATT(d|d) 
 \draw[densely dotted] (0,2.197*1.9) -- (1*1.9,2.197*1.9);

%\LACRT(d|d) discrete
%line
% \draw (0,0) -- (1*1.9,2.197*1.9);
%label
% \node[ circle,inner sep=0pt,fill=white,label= above right:{\tiny \begin{tabular}{c c}$\LACRT(d|d)$ \\({discrete})\end{tabular}}] at (1.9*1.68,1.9*.95) (D){};
%arrow
%\draw [-{Stealth[length=2mm]}] (1.9*0.57,1.9*1.178) -- (1.9*1.8,1.9*1.178);
 
%ACRT(d'|d) discrete
%line
 \draw (1*1.9,2.197*1.9) -- (4*1.9,3.497*1.9); 
%label
 \node[circle,inner sep=0pt,fill=white,label=below right:{\tiny \begin{tabular}{c c}$\LACRT(d'|d)$ \\(discrete)\end{tabular}}] at (2*2.3,1.95*2.43) (F){};
%arrow 
 \draw [-{Stealth[length=2mm]}] (1.9*2,1.9*2.63) -- (1.9*2.6,1.9*2.43);

%\LACRT(d|d) continuous
%line
 \draw[smooth] (0.28*1.9,1.57*1.9) -- (2*1.9,3.135*1.9);
%label
 \node [ circle,inner sep=0pt, fill=white,label= above :{\tiny \begin{tabular}{c c}${\LACRT(d|d)}$ \\({continuous})\end{tabular}}] at (1.9*0.8,1.9*2.93) (B){};
%arrow
\draw [-{Stealth[length=2mm]}] (1.9*1.25,1.9*2.492) -- (1.9*0.8,1.9*3);

%ACRT(d'|d) continuous
%line
 \draw[smooth] (2.35*1.9,3.135*1.9) -- (5.2*1.9,3.788*1.9);
%label
 \node[circle,inner sep=0pt,fill=black,label=above left:{\tiny \begin{tabular}{c c}$\LACRT(d'|d)$ \\ (continuous)\end{tabular}}] at (1.9*4.6,1.9*3.65) (H){};
%arrow
 \draw [-{Stealth[length=2mm]}]  (1.9*5,1.9*3.73) -- (1.9*4.5,1.9*3.81);
     
\end{tikzpicture}
    \label{fig:F2}
   \end{center}
   \justifying
   \setstretch{.75}    {\vspace{-.5cm}\scriptsize  \textit{Notes:} The figure plots $\LATT(\cdot|d)$ (the average effect of experiencing each dose for dose group $d$). We highlight causal parameters for two doses, $d$ and $d^\prime$. $\LATT(d|d)$ and $\LATT(d^\prime|d)$ are average treatment effect on the treated parameters and refer to the height of the curve. $\LACRT(d|d)$ and $\LACRT(d^\prime|d)$ are average causal response parameters and refer to the slope of the curve. We show them for a continuous dose, when the $\LACRT$ is the slope of a tangent line, and for a discrete dose when $\LACRT$ is the slope of a line connecting two discrete points on $\LATT(d|d)$.}
\end{figure}

\Cref{fig:F2} illustrates these parameters graphically. The concave line plots an average treatment effect function against the dose for dose group $d$, $\LATT(\cdot|d)$. If we consider dose levels $d$ and $d^\prime$, there are two possible $\LATT$ parameters. The first, $\LATT(d|d)$, the level of dose group $d$'s average treatment effect function at $d$, is an average treatment effect that is ``local'' to units that experienced dose $d$. The second, $\LATT(d^\prime|d)$, is also ``local'' to dose group $d$, but refers to the effect they would experience at dose $d^\prime$ even though they did not actually receive that dose. The continuous-dose $\LACRT$ parameters are the slopes of tangent lines to the $\LATT(\cdot|d)$ function, and the discrete-dose $\LACRT$ parameters are the slopes of lines connecting two points on the $\LATT(\cdot|d)$ function. As with $\LATT$'s, our definitions encompass causal responses to doses other than the one a group actually receives (i.e., $\LACRT(d^\prime|d)$).

A proper interpretation of continuous DiD results hinges on which type of parameter one wants, can identify, and can estimate. For instance, even if all $\LATT(d|d)$ parameters are large and positive, some $\LACRT(d|d)$ parameters could be zero or negative. A researcher misinterpreting a large $ATT$ estimate as an $\ACRT$, in this case, would mistakenly conclude that a policy to raise every unit's dose would have large effects. A researcher confusing a small $\ACRT$ for an $ATT$ would mistakenly conclude that a policy was ineffective, even though it actually just has small effects at the margin. 

The above-mentioned causal parameters are functional parameters because they are allowed to vary arbitrarily across dose groups $d$ and across (counterfactual) doses $d'$.  This contrasts with $\beta^{twfe}$ from \eqref{eqn:twfe}, which is a single number.  In many applications, it may be desirable to aggregate these functional parameters into lower-dimensional objects that are easier to report and may be more precisely estimated.  We focus on aggregations that average the functional parameters discussed above using the distribution of the dose among all treated units,
\vspace{-.1cm}\begin{align*}
    \LATT^{\loc} = \E[\LATT(D|D) | D>0] \qquad \textrm{and} \qquad \ATT^{\glob} = \E[\ATT(D)|D>0]\\
    \LACRT^{\loc} = \E[\LACRT(D|D) | D>0] \qquad \textrm{and} \qquad \ACRT^{\glob} = \E[\ACRT(D)|D>0].
\end{align*}
These provide natural ways to summarize the underlying parameters.  We use the $\loc$ superscript to denote that $\LATT^{\loc}$ and $\LACRT^{\loc}$ summarize treatment effects that are local effects of particular doses, while we use the superscript $\glob$ to denote that $\ATT^{\glob}$ and $\ACRT^{\glob}$ summarize treatment effects of particular doses globally (i.e., across all treated units).  All four of these parameters provide ``best'' approximations in the sense of minimizing the mean squared distance between the summary parameter and the functional parameters. 
Also, note that $\LACRT^\loc$ and $\ACRT^{\glob}$ are average derivative-type parameters, and average derivatives have been widely studied in econometrics.

\subsection{Identification with a Continuous Treatment} \label{sec:identification}
This section discusses the identification of average treatment effect and average causal response parameters. Toward this end, we make the following assumptions.

\begin{assumption}[Random Sampling] \label{ass:random-sampling} \vspace{-.5cm}\singlespacing The observed data consist of  $\{Y_{i,{t=2}},Y_{i,t=1},D_i\}_{i=1}^n$, which is independent and identically distributed.
\end{assumption}

\begin{assumption}[Treatment]\label{ass:treatment-minimal}
 \vspace{-.5cm}\singlespacing  In period $t=1$, no unit is treated, while in period $t=2$, the treatment dosage $D$ has support $\mathcal{D} = \{0\} \cup \mathcal{D}_+$, where $\mathcal{D}_+ \subseteq (0, \infty)$.  In addition, $\P(D=0) > 0$.
\end{assumption}

\begin{assumption}[No-Anticipation and Observed Outcomes] \label{ass:no-anticipation}\vspace{-.5cm}\singlespacing For all units, and all $d\in\mathcal{D}$,
    \vspace{-.1cm}\begin{align*}
    Y_{i,t=1} = Y_{i,t=1}(d) = Y_{i,t=1}(0) \quad \textrm{and} \quad Y_{i,t=2} = Y_{i,t=2}(D_i).
\end{align*}
\end{assumption}

\Cref{ass:random-sampling} says that we observe two periods of $iid$ panel data. \Cref{ass:treatment-minimal} formalizes that a mass of units do not participate in the treatment in either period (we discuss the case with no untreated units in more detail at the end of this section), and the rest receive a positive amount of the treatment that can vary in amount across units.   
\Cref{ass:no-anticipation} says that units do not anticipate future treatments, so we observe untreated potential outcomes for all units in the first period. In the second period, we observe the potential outcome corresponding to the actual dose that unit $i$ experienced.

\subsubsection{Identification under parallel trends}
Identification of average level treatment effects follows closely from the DiD setup with binary treatments. In particular, our results rely on an extension of the binary parallel trends assumption. 
\begin{namedassumption}{PT} [Parallel Trends] \label{ass:continuous-parallel-trends} \vspace{-.5cm}\singlespacing For all $d \in \mathcal{D}_+$,
  \begin{align*}
    \E[Y_{t=2}(0) - Y_{t=1}(0) | D=d] = \E[Y_{t=2}(0) - Y_{t=1}(0) | D=0].
  \end{align*}
\end{namedassumption}
\Cref{ass:continuous-parallel-trends} says that the average evolution of outcomes that units with any dose $d$ would have experienced without treatment is the same as the evolution of outcomes that units in the untreated group actually experienced. Binary DiD designs also rely on assumptions like this. To simplify the exposition below, we often simply refer to \Cref{ass:continuous-parallel-trends} as \textit{parallel trends} (PT). The following result shows that under \Cref{ass:continuous-parallel-trends}, $\LATT(d|d)$ is identified; all proofs are in \Cref{app:proofs}. 
\begin{theorem} \label{thm:att} \vspace{-.3cm}\singlespacing Under Assumptions \ref{ass:random-sampling}, \ref{ass:treatment-minimal}, \ref{ass:no-anticipation}, and \ref{ass:continuous-parallel-trends}, $\LATT(d|d)$ is identified for all $d \in \mathcal{D}_+$, and it is given by
\vspace{-.1cm}\begin{align*}
    \LATT(d|d) = \E[\Delta Y | D=d] - \E[\Delta Y | D=0].
\end{align*}
Furthermore, $\LATT^{\loc} = \E[\Delta Y | D>0] - \E[\Delta Y | D=0]$.
\end{theorem}
\Cref{thm:att} states that $\LATT(d|d)$ equals the difference between the change in outcomes for dose group $d$ and the untreated group. It generalizes \citet{fricke-2017}'s result for two doses to richer treatment patterns. As a direct consequence, by averaging all the $\LATT(d|d)$'s over the distribution of non-zero dosages, we have that the summary parameter $\LATT^{\loc}$ is identified by simply comparing units with a positive dose to untreated units. On the other hand, parallel trends, as defined in \Cref{ass:continuous-parallel-trends}, is \emph{not} strong enough to guarantee the identification of $\ATT(d)$.

The identification of average causal response parameters differs from the identification of $\LATT$ parameters because it requires comparisons between dose groups.  
\begin{assumption}[Continuous or Multi-Valued Discrete Treatment]\label{ass:treatment}
The treatment is either continuous or multi-valued discrete.  More precisely, one of the following is true:
\begin{itemize}
    \item[(a)] $\mathcal{D}_+ =\mathcal{D}_+^c$, where $\mathcal{D}_+^c = (d_L, d_U)$  with $f_{D|D>0}$ a Lebesgue density which satisfies $f_{D|D>0}(d) > 0$ for all $d \in \mathcal{D}_+^c$, and $\E[\Delta Y|D=d]$ is differentiable on $\mathcal{D}_+^c$.
    \item[(b)] $\mathcal{D}_+ = \mathcal{D}_+^{mv}$ where $\mathcal{D}_+^{mv} \subset \mathbb{N}_+$, with $\mathbb{N}_+ = \{1,2,3,\ldots\}$ denotes the strictly positive natural numbers.  Let $d_j$ denote the $j^{\text{th}}$ element of $\mathcal{D}_+^{mv}$.  In addition, $\P(D=d) > 0$ for all $d \in \mathcal{D}_+^{mv}$.
\end{itemize}
\end{assumption}
\Cref{ass:treatment} distinguishes between cases with a continuous \ref{ass:treatment}(a) or discrete  \ref{ass:treatment}(b) treatment.  \Cref{ass:treatment}(a) allows for the smallest value of the treatment to be arbitrarily close to zero or strictly larger than zero, both of which are common in applications.

Our central identification result is that causal response parameters are not identified under \Cref{ass:continuous-parallel-trends}, because comparisons between different dose groups are biased when treatment effects (of the same dose) vary across dose groups, even when the average evolution of untreated potential outcomes is the same.

\begin{theorem} \label{thm:acr} \vspace{-.3cm}\singlespacing Under Assumptions \ref{ass:random-sampling}, \ref{ass:treatment-minimal}, \ref{ass:no-anticipation}, and \ref{ass:continuous-parallel-trends}, comparisons of paths of outcomes among different dose groups recover a mix of causal effect parameters and selection bias terms. Specifically, 
\begin{itemize}
    \item[(a)] \vspace{-.1cm}For $(h,l) \in \mathcal{D}_+ \times \mathcal{D}_+$,
    \begin{align*}
        \E[\Delta Y | D=h] - \E[\Delta Y | D=l] &= \LATT(h|h) - \LATT(l|l) \\
        &= \underbrace{\E[Y_{t=2}(h) - Y_{t=2}(l) | D=h]}_{\textrm{causal effect}} + \underbrace{\Big(\LATT(l|h) - \LATT(l|l) \Big)}_{\textrm{selection bias}}.
    \end{align*}
    \item[(b)] \vspace{-.5cm} If \Cref{ass:treatment}(a) also holds, then, for $d \in \mathcal{D}_+^c$,
    \begin{align*}
            \frac{\partial \E[\Delta Y | D=d]}{\partial d} = \frac{\partial \LATT(d|d)}{\partial d}  &= \LACRT(d|d) + \underbrace{\frac{\partial \LATT(d|l)}{\partial l} \Big|_{l=d}}_{\textrm{local selection bias}}; 
    \end{align*}
    \item[(c)] \vspace{-.5cm}Alternatively, if \Cref{ass:treatment}(b) also holds, %taking $h=d_j$ and $l=d_{j-1}$ implies that
    \begin{align*}
        \frac{\E[\Delta Y | D=d_j] - \E[\Delta Y | D=d_{j-1}]}{d_j - d_{j-1}} %&= \LATT(d_j|d_j)-\LATT(d_{j-1}|d_{j-1}) \\      
        &= \LACRT(d_j|d_j) + \underbrace{\frac{\LATT(d_{j-1}|d_j) - \LATT(d_{j-1}|d_{j-1})}{d_j-d_{j-1}}}_{\textrm{scaled selection bias}}.
    \end{align*}
\end{itemize}
\end{theorem}

\Cref{thm:acr} says that under parallel trends, comparisons of outcome paths between higher- and lower-dose groups mix together (i) causal responses and (ii) a ``selection bias'' type of term that comes from differences in average treatment effects of the same dose for different dose groups.  Intuitively, even if untreated potential outcomes evolve in the same way, observed paths of outcomes differ between dose groups for two reasons. One is the causal response itself, which comes from differences in doses ($h$ versus $l$) causing differences in outcomes. The other is a selection bias type of contamination, which comes from differences across dose groups in the average level effect of the particular dose $l$---parallel trends does not rule out that different dose groups could experience different treatment effects of the same dose.  

\begin{figure}[ht]
   \begin{center}
   \caption{{Non-Identification of Average Causal Response with Treatment Effect Heterogeneity, Two Discrete Doses}}
\begin{tikzpicture}[scale=.7]
%x-axis
  \draw[->] (0, 0) -- (10, 0) node[right] {\scriptsize $l$};
%y-axis (take out label)
  \draw[->] (0, 0) -- (0, 7);
  %{$Y_{t}(d)-Y_{t}(0)$}
%axis labels
  \node[circle,inner sep=0pt,fill=black,label=left:{\tiny $\LATT(d'|d')$}] at (0,3.497*1.5) {};
  \node[circle,inner sep=0pt,fill=black,label=left:{\tiny $\LATT(d|d')$}] at (0,2.197*1.5) {};
  \node[circle,inner sep=0pt,fill=black,label=left:{\tiny $\LATT(d|d)$}] at (0,0.693*1.5) {};
  \node[circle,inner sep=0pt,fill=black,label=below:{\tiny $d'=d+1$}] at (4*1.5,0) {};
  \node[circle,inner sep=0pt,fill=black,label=below:{\tiny $d$}] at (1*1.5,0) {};
  
%curves--kind of cool how you define the domain, the variable name, and the function!
  \draw[scale=1.5, domain=0:5, smooth, variable=\d, black] plot ({\d}, {ln(1+\d)});
  \draw[scale=1.5, domain=0:5, smooth, variable=\d, black] plot ({\d}, {ln(1+8*\d)});
%label for lower treatment effect group
   \node[circle,inner sep=0pt,fill=white,label=below right:{\tiny $\LATT(l|d)$}] at (5*1.5,1.7*1.75) {};
%label for higher treatment effect group
   \node[circle,inner sep=0pt,label=below right:{\tiny $\LATT(l|d')$}] at (4*1.9,4.1*1.45) {};

%ACRT(d'|d') line
  \draw[line width=.3mm] (1*1.5,2.197*1.5) -- (4*1.5,3.497*1.5);
%l-shape on the ACRT line
    \draw (1.5*1.5, 2.41*1.5) -- (1.5*1.5,2.7*1.5);
    \draw (2.18*1.5, 2.7*1.5) -- (1.5*1.5,2.7*1.5);
%label for ACRT
    \node[circle,inner sep=0pt,fill=white,label=above left:{\tiny  $\LACRT(d'|d')$}]  at (1.5*1.6,2.7*1.5) {};
    
%Estimator line
    \draw[line width=.3mm] (1*1.5, 0.693*1.5)-- (4*1.5,3.497*1.5);
%l-shape on the estimator line
  \draw (3*1.5, 2.56*1.5) -- (3.5*1.5,2.56*1.5);
  \draw (3.5*1.5, 2.56*1.5) -- (3.5*1.5,3.027*1.5);
%right endpoint of estimator
   \node[circle,draw,line width=1pt,inner sep=0pt,minimum size=6pt,label=below right:{}] at (4*1.5,3.497*1.5) {};
%left endpoint of estimator
    \node[circle,draw,line width=1pt,inner sep=0pt,minimum size=6pt,label=below right:{}] at (1*1.5,0.693*1.5) {};
%label for the estimator (ACRT + bias)
\node[circle,inner sep=0pt,fill=white,label=right:{ \tiny \begin{tabular}{l c} $=\LACRT(d'|d')+ \LATT(d|d') -\LATT(d|d)$ \end{tabular}}]  at (1.5*3.3,2.66*1.5) {};

%vertical and horizontal lines for the two ATT points
   \draw[densely dotted] (4*1.5,0) -- (4*1.5,3.497*1.5);
   \draw[densely dotted] (0,1.5*3.497) -- (4*1.5,3.497*1.5);
   \draw[densely dotted] (1*1.5,0) -- (1*1.5,0.693*1.5);
   \draw[densely dotted] (0,0.693*1.5) -- (1*1.5,0.693*1.5);

%vertical and horizontal line for \LATT(d|d')
   \draw[densely dotted] (1*1.5, 0.693*1.5)-- (1*1.5,2.197*1.5);
   \draw[densely dotted] (0, 2.197*1.5)-- (1*1.5,2.197*1.5);

\end{tikzpicture}
    \label{fig:F3}
   \end{center}
   \justifying
   \setstretch{.75}
   {\vspace{-.5cm}\scriptsize  \textit{Notes:} The figure shows that comparing adjacent $\LATT(d|d)$'s equals an $\LACRT$ parameter (the slope of the higher-dose group's $\LATT$ function) and selection bias (the difference between the two groups' $\LATT$ functions at the lower dose).}
    %\end{flushleft}
\end{figure}

Figure \ref{fig:F3} illustrates this result for an example with two dose groups and two doses: $d$ and $d^\prime = d+1$.  The slope of the line that connects the points $(d,\LATT(d|d))$ and $(d^\prime,\LATT(d^\prime|d^\prime))$ is steeper than the average causal response of interest, $\LACRT(d^\prime |d^\prime )$, because it jumps from one $\LATT$ function to the other. %\footnote{Because we are considering one unit increments, the bias can be seen on the $y$-axis as well: $\LATT(d|d^\prime)-\LATT(d|d)$ is bias and $\LATT(d^\prime|d^\prime)-\LATT(d|d^\prime)$ is the $\LACRT(d^\prime|d^\prime)$.} 
This is captured by the selection bias term, a version of selection-on-gains that equals the difference in treatment effects at the lower dose: $\LATT(d|d^\prime)-\LATT(d|d)$. It breaks the causal interpretation because observed outcomes for lower-dose units are not a valid counterfactual for what higher-dose units would have experienced at the lower dose. The selection bias is not identified as we do not observe $Y_{t=2}(d)$ for units that experienced dose $d^\prime$. Such a result precludes a causal interpretation of $\LATT(d|d)$ differences across doses under Assumption PT.

\subsubsection{Identification under strong parallel trends}
 This section discusses an alternative, typically stronger assumption that allows for the identification of $\ACRT(d)$ and $\ATT(d)$ parameters, which we refer to as \textit{strong parallel trends} (SPT). %Henceforth, we use strong parallel trends and SPT to refer to \Cref{ass:strong-parallel-trends} and use these as synonymous.
\begin{namedassumption}{SPT} [Strong Parallel Trends] \label{ass:strong-parallel-trends} \vspace{-.3cm}\singlespacing For all $d \in \mathcal{D}$,
  \begin{align*}
    \E[Y_{t=2}(d) - Y_{t=1}(0)|D>0] = \E[Y_{t=2}(d) - Y_{t=1}(0) | D=d].
  \end{align*}
\end{namedassumption}
Under \Cref{ass:no-anticipation}, the right-hand side of the equation in \Cref{ass:strong-parallel-trends} is the (observed) average evolution of outcomes for dose group $d$.  \Cref{ass:strong-parallel-trends} says that the average evolution of outcomes for the entire treated population if all experienced dose $d$ (the left-hand side of the previous equation) is equal to the path of outcomes that dose group $d$ actually experienced.  In applications where the treatment is binary, \Cref{ass:strong-parallel-trends}, like \Cref{ass:continuous-parallel-trends}, reduces to the usual parallel trends assumption.  Like the case with a binary treatment, it allows for treated units to select into being treated.  Among treated units, though, it rules out selection into a particular amount of the treatment.  With more complicated treatments, \Cref{ass:strong-parallel-trends} notably differs from \Cref{ass:continuous-parallel-trends} because it involves potential outcomes under different doses, $Y_t(d)$, rather than only untreated potential outcomes, $Y_t(0)$.  While \Cref{ass:strong-parallel-trends} is not strictly stronger than \Cref{ass:continuous-parallel-trends} (e.g., notice that it does not require parallel trends in untreated potential outcomes for all dose groups), we refer to it as \textit{strong parallel trends} to indicate that in many applications it would be a stronger, perhaps much stronger, assumption. 

An alternative way to think about \Cref{ass:strong-parallel-trends} is as an assumption that restricts treatment effect heterogeneity.\footnote{There are some instances of versions of strong parallel trends implicitly being discussed in empirical work. \citet[p.\,1636]{chodorow-reich-nenov-simsek-2021}'s cross-region study of marginal propensities to consume (MPC) notes the possibility of finding a zero even when the MPC>0 in all areas: ``if low wealth areas have high MPCs and high wealth areas have low MPCs, an increase in the stock market could induce the same change in spending in both low and high wealth areas.''
Similarly, \citet[p.\,25]{saez-slemrod-giertz-2012} discuss a more restrictive version of strong parallel trends in the context of estimating the elasticity of taxable income for two groups facing different positive tax changes: ``if the control group faces a tax change, difference-in-differences estimates will be consistent only if the elasticities are the same for the two groups.'' \label{fn:spt-in-empirical-work}} In particular, if one maintains \Cref{ass:continuous-parallel-trends}, \Cref{ass:strong-parallel-trends} is equivalent to assuming that $\LATT(d|d) = \ATT(d)$ for all doses. This condition can also be viewed as a structural assumption in the sense that it effectively allows one to extrapolate treatment effects of dose $d$ among dose group $d$ to treatment effects of dose $d$ for the entire treated population.

In the remainder of this section, we show that \Cref{ass:strong-parallel-trends} is useful for recovering ``global'' average causal effect parameters, which are straightforward to compare to each other, and, hence, sidestep the selection bias issues discussed above.  Before doing that, it is worth mentioning that we are not proposing \Cref{ass:strong-parallel-trends} as an assumption that empirical researchers should readily adopt; in fact, in many applications, \Cref{ass:strong-parallel-trends} may be a strong or implausible assumption.  Rather, our aim is to clarify that many natural target parameters in DiD applications with a continuous treatment require stronger assumptions than the parallel trends as defined in \Cref{ass:continuous-parallel-trends}.

\begin{theorem} \label{thm:ate} \vspace{-.3cm}\singlespacing Under Assumptions \ref{ass:random-sampling}, \ref{ass:treatment-minimal}, \ref{ass:no-anticipation}, and \ref{ass:strong-parallel-trends},
\begin{itemize}
    \item[(a)] \vspace{-.2cm} For $d \in \mathcal{D}_+$, it follows that
    \begin{align*}
        \ATT(d) = \E[\Delta Y | D=d] - \E[\Delta Y | D=0].
    \end{align*}
    \item[(b)] \vspace{-.1cm} For $(h,l) \in \mathcal{D}_+ \times \mathcal{D}_+$,
    \begin{align*}
        \E[Y_{t=2}(h) - Y_{t=2}(l)|D>0] = \ATT(h) - \ATT(l) =  \E[\Delta Y | D=h] - \E[\Delta Y | D=l] 
    \end{align*}
    \item[(c)] \vspace{-.1cm}When \Cref{ass:treatment}(a) holds (i.e., treatment is continuous), it follows that, for $d \in \mathcal{D}_+^c$,
    \begin{align*}
             \ACRT(d) = \frac{\partial \E[\Delta Y | D=d]}{\partial d} 
    \end{align*}
    \item[(d)] \vspace{-.1cm}When \Cref{ass:treatment}(b) holds (i.e., treatment is discrete), it follows that 
    \begin{align*}
          \ACRT(d_j) = \frac{\E[\Delta Y | D=d_j] - \E[\Delta Y | D=d_{j-1}]}{d_j - d_{j-1}} 
    \end{align*}
\end{itemize}
\end{theorem}
For part (a) of \Cref{thm:ate}, recall that $\LATT(d|d)$ and $\ATT(d)$ differ when there is selection into dose group $d$ on the basis of treatment effects. Strong parallel trends rules out that kind of selection, which means that comparing average outcome changes of dose group $d$ to the untreated group identifies $\ATT(d)$.  Part (b) says that comparisons of the average change in outcomes over time for different dose groups have a causal interpretation under \Cref{ass:strong-parallel-trends}. 
For parts (c) and (d), strong parallel trends ensures that each dose group $d$ serves as a valid counterfactual for the entire treated population under that specific dose $d$, and, hence, that causal response parameters are identified under \Cref{ass:strong-parallel-trends}.

Strong parallel trends only changes the interpretation of the estimand, not its form. One important implication is that conventional pre-tests for differential changes across groups before treatment cannot distinguish between \Cref{ass:continuous-parallel-trends} and \Cref{ass:strong-parallel-trends}. That is, because only untreated potential outcomes are observed before treatment under \Cref{ass:no-anticipation}, these periods cannot test the additional content of an assumption like SPT that necessarily involves treated potential outcomes.

Finally, the identification results in \Cref{thm:ate} immediately imply that averages of the $\ATT(d)$ and $\ACRT(d)$ building blocks are identified as well. The following corollary states these results.
\begin{corollary} \label{cor:spt-agg} \vspace{-.5cm}\singlespacing Under Assumptions \ref{ass:random-sampling}, \ref{ass:treatment-minimal}, \ref{ass:no-anticipation}, and \ref{ass:strong-parallel-trends},
\begin{itemize}
    \vspace{-.1cm}\item[(a)] It follows that
    \vspace{-.1cm}\begin{align*}
    \ATT^{\glob} = \E[\Delta Y | D>0] - \E[\Delta Y | D=0].
\end{align*}
\item[(b)] \vspace{-.2cm}When \Cref{ass:treatment}(a) holds (i.e., treatment is continuous), it follows that
    \begin{align*}
             \ACRT^{\glob} = \E\left[\left.\frac{\partial \E[\Delta Y | D=d]}{\partial d}\bigg|_{d=D} \right| D>0\right] = \int_{d_L}^{d_U}\frac{\partial \E[\Delta Y | D=d]}{\partial d}\bigg|_{d=s} f_{D|D>0}(s)ds. 
    \end{align*} 
\item[(c)] \vspace{-.1cm}When \Cref{ass:treatment}(b) holds (i.e., treatment is multi-valued), it follows that
   \vspace{-.2cm}\begin{align*}
          \ACRT^{\glob} = \sum_{j=1}^{J}\left(\frac{\E[\Delta Y | D=d_j] - \E[\Delta Y | D=d_{j-1}]}{d_j - d_{j-1}}\right)\P(D=d_j|D>0).
    \end{align*}
 \end{itemize}
\end{corollary}
These results highlight how identification in continuous DiD designs is fundamentally a question about dose-specific building block parameters and the underlying parallel trends assumption, not the aggregation choices that lead to particular summary parameters. 

\begin{remark}[No untreated units]\label{rem:no_untreated_units}
    Researchers often use continuous designs when all units in their sample receive some amount of the treatment, having in mind comparing units that are ``more treated'' to units that are ``less treated''.  Without untreated units, it is infeasible to compare dose group $d$ to an untreated group, and, hence, it is infeasible to directly recover $\LATT(d|d)$ or $\ATT(d)$. However, a natural alternative is to compare dose group $d$ to dose group $d_L$ (the lowest possible amount of the treatment).  In \Cref{app:no-untreated-units} in the Supplementary Appendix, we show that, under parallel trends, when there are no untreated units, 
    \vspace{-.1cm}\begin{align*}
        \E[\Delta Y | D=d] - \E[\Delta Y | D=d_L] = \LATT(d|d) - \LATT(d_L|d_L).
    \end{align*}
    This shows that this comparison is related to underlying causal effect parameters under parallel trends; however, recall from \Cref{thm:acr} that the expression on the right-hand side mixes together the average causal response of moving from $d_L$ to $d$ with selection bias.  Under strong parallel trends, we have instead that
    \vspace{-.2cm}\begin{align*}
        \E[\Delta Y | D=d] - \E[\Delta Y | D=d_L] = \ATT(d) - \ATT(d_L) = \E[Y_{t=2}(d) - Y_{t=2}(d_L)|D>0],
    \end{align*}
    which does not include selection bias terms.  This discussion highlights that (unlike a setting with a binary treatment) continuous variation in the dose can be used to learn about causal effects even if there is no untreated comparison group, but interpreting these results as causal effects of the treatments requires strengthening \Cref{ass:continuous-parallel-trends}. See also \citet{fricke-2017} for a related discussion.
\end{remark}

\subsection{What Parameter Does TWFE Estimate?} \label{sec:twfe}

Empirical researchers using continuous DiD designs typically estimate a single summary parameter using a linear TWFE regression like \Cref{eqn:twfe}. This section links the TWFE estimand to our identification results for dose-specific parameters, describes the assumptions necessary to give TWFE \emph{some} causal interpretation, and discusses what that interpretation is.  We focus on continuous treatments and defer the discussion of multi-valued discrete treatments to \Cref{app:twfe-multivalued-treatment} in the Supplementary Appendix.

Our impression is that empirical researchers typically interpret $\beta^{twfe}$ in three main (and related) ways, implicitly relying on different building blocks.  First, $\beta^{twfe}$ is often directly interpreted as a causal response parameter; that is, how much the outcome causally increases on average when the treatment increases by one unit.  This is the causal version of how regression coefficients are often taught to be interpreted in introductory econometrics classes.  Second, it is common to pick a representative value for $d$, to report $d \times \beta^{twfe}$, and interpret this quantity as $\ATT(d)$.  This is the main interpretation provided in \citet{acemoglu-finkelstein-2008}:  ``Given that the average hospital has a 38 percent Medicare share prior to PPS, this estimate [i.e., of $\beta^{twfe}$, here equal to 1.129] suggests that in its first 3 years, the introduction of PPS was associated with an increase in the depreciation share of about 0.42 ($\approx$ 1.129 $\times$ 0.38) for the average hospital.''  Rearranging this expression shows that under this interpretation $\beta^{twfe} = \LATT(d|d)/d$, which relates $\beta^{twfe}$ to a scaled level effect.  Third, it is common to take two different representative values of the dose, $d_1$ and $d_2$---a common choice is the 25th percentiles and 75th percentiles of the dose---and interpret $\beta^{twfe}$ as the average causal response of moving from dose $d_1$ to dose $d_2$ scaled by the distance between $d_1$ and $d_2$; this is a scaled $2\times 2$ effect. We aim to assess whether such types of interpretations are justified and under which conditions.

\begin{table}[ht]
    \centering
    \caption{TWFE Decomposition Weights}
    \label{tab:twfe-weights}
    \small
     \resizebox{0.8\textwidth}{!}{%
    \setlength{\tabcolsep}{1.5pt}
    \begin{tabularx}{\textwidth}{r@{\hspace{15pt}}rclrcl}
    \toprule
     Decomposition & \multicolumn{3}{c}{ $D>0$ Weights} & \multicolumn{3}{c}{$D=0$ Weights} \\
    \midrule
    \addlinespace[10pt]
    Causal response & $w^{\textrm{acrt}}_1(l)$ & $=$  & $\displaystyle \frac{(\E[D | D \geq l] - \E[D]) \P(D \geq l)}{\Var(D)}$ & $w^{\textrm{acrt}}_0$ & $=$  & $\displaystyle \frac{(\E[D|D>0] - \E[D])\P(D>0)d_L}{\Var(D)}$ \\[10pt]
    Levels & $w_1^{\textrm{lev}}(l)$ & $=$  & $\displaystyle \frac{ (l-\E[D]) }{\Var(D)} f_{D}(l)$ & $w_0^{lev}$ & $=$ & $\displaystyle -\frac{\E[D]\P(D=0)}{\Var(D)}$ \\[10pt]
    Scaled levels & $w^{\textrm{s}}(l)$ & $=$  & $ l \displaystyle \frac{ (l-\E[D]) }{\Var(D)} f_{D}(l)$ & & & \\[10pt]
    Scaled $2 \times 2$ & $w_1^{2\times2}(l,h)$ & $=$  & $\displaystyle \frac{(h-l)^2 f_D(h) f_D(l)}{\Var(D)}$ & $w_0^{2\times2}(h)$ & $=$  & $\displaystyle \frac{h^2 f_D(h) \P(D=0)}{\Var(D)}$ \\[5pt]
    \bottomrule
    \end{tabularx}
    }
    \smallskip
    
    \RaggedRight\noindent{\scriptsize \textit{Notes:} The table provides the formulas for the weights used in the decompositions of $\beta^{twfe}$ provided in this section.}
\end{table}

The next proposition presents our decompositions of $\beta^{twfe}$ under parallel trends (\Cref{ass:continuous-parallel-trends}) and under strong parallel trends (\Cref{ass:strong-parallel-trends}). 
 The decompositions differ on the basis of the underlying building block parameters: causal response parameters ($\LACRT(d|d)$ and $\ACRT(d)$), level treatment effect parameters ($\LATT(d|d)$ and $\ATT(d)$), scaled level effects ($\LATT(d|d)/d$ and $\ATT(d)/d$), or scaled $2 \times 2$ effects ($\E[Y_{t=2}(h)-Y_{t=2}(l)|D=h]/(h-l)$ and $\E[Y_{t=2}(h)-Y_{t=2}(l)|D>0]/(h-l)$). These building blocks are connected with the dose-parameters discussed in Section \ref{sec:identification} and how empirical researchers interpret  $\beta^{twfe}$; see \Cref{app:additional-twfe-decomposition-results} in the Supplementary Appendix for additional decompositions based on different building blocks.  The weights attached to each of these decompositions are presented in Table \ref{tab:twfe-weights}.  

\begin{theorem} \label{thm:twfe-pt} \vspace{-.5cm}\singlespacing Under Assumptions \ref{ass:random-sampling}, \ref{ass:treatment-minimal}, \ref{ass:no-anticipation}, \ref{ass:treatment}(a), and \ref{ass:continuous-parallel-trends},  $\beta^{twfe}$ can be decomposed in the following ways:
    \begin{itemize}\vspace{-.2cm}
    \item[(a)] Causal Response Decomposition:\vspace{-.1cm}
    \begin{align*}
        \beta^{twfe} &= \int_{d_L}^{d_U} w^{acrt}_1(l) \left(\LACRT(l|l) + \underbrace{\frac{\partial \LATT(l|h)}{\partial h}\Big|_{h=l}}_{\textrm{selection bias}} \right) \, dl + w^{acrt}_0 \frac{\LATT(d_L|d_L)}{d_L} %+ \underbrace{\int_{d_L}^{d_U} w^{acrt}_1(l) \frac{\partial \LATT(l|h)}{\partial h}\Big|_{h=l} \, dl}_{\textrm{selection bias}},
    \end{align*}
    where the weights are always positive and integrate to 1.
    \item[(b)] \vspace{-.2cm}Levels Decomposition:\vspace{-.4cm}
    \begin{align*}
        \beta^{twfe} = \int_{d_L}^{d_U}  w_1^{\textrm{lev}}(l) \LATT(l|l) \, dl,
    \end{align*}   
    where $w_1^{\textrm{lev}}(l)\lessgtr 0$ for $l\lessgtr \E[D]$, and $\int_{d_L}^{d_U}  w_1^{\textrm{lev}}(l)\, dl + w_0^{lev} = 0$.
    \item[(c)]\vspace{-.2cm} Scaled Levels Decomposition:\vspace{-.4cm}
    \begin{align*}
        \beta^{twfe} = \int_{d_L}^{d_U}  w^{\textrm{s}}(l) \frac{\LATT(l|l)}{l} \, dl ,
    \end{align*}
    where $w^{\textrm{s}}(l)\lessgtr 0$ for $l\lessgtr \E[D]$, and $\int_{d_L}^{d_U}  w^{\textrm{s}}(l)\, dl = 1$.
    \item[(d)] Scaled $2\times 2$ Decomposition
    \begin{align*}
    \beta^{twfe} &= \int_{d_L}^{d_U} \int_{\mathcal{D}, h>l} w_1^{2\times2}(l,h) \left(\underbrace{\frac{\E[Y_{t=2}(h) - Y_{t=2}(l) | D=h]}{h-l}}_{\textrm{causal response}} +\underbrace{\frac{\LATT(l|h) - \LATT(l|l)}{h-l}}_{\textrm{selection bias}}\right)\, dh \, dl \\
    & \quad \quad +\int_{d_L}^{d_U} w_0^{2\times2}(l) \frac{\LATT(l|l)}{l} \, dl,
    \end{align*}
    where %$w_1^{2\times2}(l,h) = (h-l)^2 f_D(h) f_D(l) / \Var(D)$, $w_0^{2\times2}(h) = h^2 f_D(h) p_0^D / \Var(D)$.  
    the weights $w_1^{2\times2}$ and $w_0^{2\times2}$ are always positive and integrate to 1.
\end{itemize}
    \onehalfspacing If one imposes \Cref{ass:strong-parallel-trends} instead of \Cref{ass:continuous-parallel-trends}, then the selection bias terms from Part (a) and Part (d) become zero, and the remainder of the decompositions remain true, except one needs to replace $\LACRT(l|l)$ with $\ACRT(l)$ in Part (a), $\LATT(l|l)$ with $\ATT(l)$ in Parts (b), (c) and (d), and $\E[Y_{t=2}(h) - Y_{t=2}(l) | D=h]$ with $\E[Y_{t=2}(h) - Y_{t=2}(l)|D>0]$ in Part (d).
\end{theorem}

\Cref{thm:twfe-pt} shows that the same TWFE estimand yields very different decomposition results, depending on the type of parallel trends used and the particular causal parameter employed as a building block for the analysis. Despite these multiple possible decompositions, one important feature that arises from \Cref{thm:twfe-pt} is that the weighting associated with any of the decompositions of $\beta^{twfe}$ has some undesirable properties, making $\beta^{twfe}$ an unappealing causal summary parameter in DiD setups with continuous treatments. Yet, each of these different decompositions highlights distinct concerns, as we discuss below.

\Cref{thm:twfe-pt}(a) shows that when causal responses are taken as the building blocks of the analysis, under \Cref{ass:continuous-parallel-trends}, $\beta^{twfe}$ is equal to a convex weighted average of $\LACRT(d|d)$ and the same selection bias derived in \Cref{thm:acr}.\footnote{Part (a) is mechanically related to the results in \citet{yitzhaki-1996} on interpreting linear projection coefficients with a continuous regressor when the conditional expectation may be nonlinear. Part (a) also includes a term that shows how TWFE handles a discrete jump from 0 to the minimum treated dose, $d_L$. Paths of outcomes are not observed for doses below $d_L$, but the scaled $ATT$ for dose group $d_L$, $\LATT(d_L|d_L)/d_L$, is averaged into $\beta^{twfe}$.\label{fn:yitzhaki}}  The sign of this selection bias depends on how treatment effects vary across dose groups at a given dose. If units in higher dose groups would have had larger positive treatment effects at every dose, for example, then $\beta^{twfe}$ will be larger than the weighted average of the $\LACRT(d|d)$'s that appear in \Cref{thm:twfe-pt}(a). \Cref{fig:F3} illustrates this case for two groups. Strong parallel trends eliminates the selection bias term, but does not affect the weights.

Even under strong parallel trends, the particular interpretation of $\beta^{twfe}$ in terms of $\ACRT(d)$'s hinges on the aggregation embodied in the weights $w^{acrt}_1(d)$.  Because $w^{acrt}_1(d)$ is positive and integrates to 1, under \Cref{ass:strong-parallel-trends} $\beta^{twfe}$ is \textit{weakly causal} \citep{blandhol-bonney-mogstad-torgovitsky-2025}. However, it does not estimate a natural target parameter like $\ACRT^{\glob}$ because the TWFE weights do not generally equal the dose distribution among treated, $f_{D|D>0}(d)$. Interestingly, the weights $w^{acrt}_1(l)$ underlying $\beta^{twfe}$ depend on the entire distribution of the dose, making it sensitive to the size of the untreated group. This property is rather unappealing.  For example, in our application, if we drop the untreated group (dropping the untreated group does not alter the underlying average causal responses), our estimate of $\beta^{twfe}$ shrinks by 78\%.  Instead of letting the estimation method implicitly summarize the $\ACRT$'s, we recommend that researchers choose these aggregation schemes explicitly. In our view, a natural and econometrically-guided way to aggregate the $\ACRT$'s into a summary parameter is given by $\ACRT^{\glob}$, which is identified (as indicated in \Cref{cor:spt-agg}) and can also be easily estimated.

Part (b) expresses $\beta^{twfe}$ as a weighted integral of $\LATT(d|d)$ under parallel trends with weights that integrate to zero rather than one. Therefore, some weights are negative, and, hence, $\beta^{twfe}$ is not weakly causal when $\LATT(d|d)$ is taken as the building block.  More significantly, $\beta^{twfe}$ puts the same amount of negative weight on $\LATT(d|d)$'s for doses below $\E[D]$ as it does positive weight on $\LATT(d|d)$'s for doses above $\E[D]$. One way to view this result is that TWFE uses above-average dose units as an ``effective treated group'' and below-average dose units as an ``effective comparison group'' that potentially includes some treated units. While the cumulative positive weights and negative weights are equal to each other, they do not generally integrate to one within these groups, which means that $\beta^{twfe}$ does not equal the difference between a weighted average of outcome paths for the effective treated group relative to the effective comparison group. 
In \Cref{app:additional-twfe-decomposition-results} in the Supplementary Appendix, however, we bridge this gap and derive a corollary of the result in Part (b) that makes the scaling issue related to this interpretation explicit and allows us to express $\beta^{twfe}$ as the following weighted Wald-estimand:
\begin{align} \label{eq:twfe-binarized}
    \beta^{twfe} = \frac{\E\Big[w^{bin}_1(D) \Delta Y \Big| D > \E[D]\Big] - \E\Big[w^{bin}_0(D) \Delta Y \Big| D < \E[D]\Big]}{\E\Big[w^{bin}_1(D) D \Big| D > \E[D]\Big] - \E\Big[w^{bin}_0(D) D \Big| D < \E[D]\Big]}.
\end{align}

\noindent The numerator of Equation \eqref{eq:twfe-binarized} shows that $\beta^{twfe}$ compares weighted average outcome changes above and below $\E[D]$ with weights proportional to how far a unit's dose is from $\E[D]$.\footnote{The exact expressions for the weights are  $w^{bin}_1(d) = \frac{|d-\E[D]|}{\E\big[|D-\E[D]|\big| D > \E[D] \big]}$ and $w^{bin}_0(d) = \frac{|d-\E[D]|}{\E\big[|D-\E[D]|\big| D \leq \E[D]\big]}$. See \Cref{app:additional-twfe-decomposition-results} in the Supplementary Appendix for more details.}  The denominator scales this comparison by the same weighted difference in $D$. This representation highlights some challenges of using $\beta^{twfe}$ to summarize the average level-effect of a continuous treatment.  First, while the numerator is (roughly) a weighted level-effect, the denominator shows that $\beta^{twfe}$ additionally depends on a measure of the average distance between the effective treated and comparison group. Second, the effective comparison group can include treated units.  %leading to negative weights on some $ATT$ parameters.  
Third, $\beta^{twfe}$ uses ``distance'' weights $w^{bin}$'s to aggregate across dosages. In contrast, $\LATT^{\loc}$ does not suffer from any of these issues.  In applications where the researcher is targeting level-effect parameters, we recommend favoring $\LATT^{\loc}$ vis-a-vis $\beta^{twfe}$.

Parts (c) and (d) of \Cref{thm:twfe-pt} provide interpretations of $\beta^{twfe}$ taking scaled paths of outcomes as building blocks.  For part (c), $\LATT(d|d)/d$ (under parallel trends) and $\ATT(d)/d$ (under strong parallel trends) are ``per-dosage'' causal parameters.  This part shows that the TWFE estimand includes negative weights under the same conditions as in part (b), though the weights integrate to one. We note that, in the case of a discrete dose, this result in part (c) corresponds to the one in Theorem S3 of the Supplementary Appendix of \citet{chaisemartin-dhaultfoeuille-2020}. Therefore, using ``average slopes'' as the underlying parameter of interest eliminates neither TWFE's potential for negative weights nor its non-intuitive weighting scheme.  For part (d), when $\beta^{twfe}$ is interpreted in terms of all possible $2 \times 2$ comparisons of changes of outcomes for higher dose groups relative to lower dose groups, the weights are all positive and integrate to 1, but, under parallel trends, these comparisons all mix causal effects of the higher treatment with selection bias terms. Although strong parallel trends removes the selection bias, the weights attached to the causal parameters are still hard to interpret.

\begin{remark}[Decomposition with no untreated units]
    It is straightforward to extend the TWFE decompositions discussed above to settings with no untreated units.  For the causal response decomposition (part (a)), the exact same result applies, with the exception that the second term involving $w_0^{acrt}$ is equal to 0.  Similarly, for the scaled $2 \times 2$ decomposition (part (d)), nothing changes except that the second term involving $w_0^{2 \times 2}$ is equal to 0.  For the levels decomposition and the scaled levels decomposition (parts (b) and (c)), with no untreated units, $\LATT(d|d)$ (or $\ATT(d)$) is not identified; instead, along the lines mentioned in \Cref{rem:no_untreated_units}, instead of using the untreated comparison group, we can instead compare to the path of outcomes of the ``least treated''.  Thus, the same decompositions continue to apply except that $\LATT(l|l)$ should be replaced by $\LATT(l|l) - \LATT(d_L|d_L)$.  This immediately means that these decompositions (in addition to negative weights) become complicated by issues related to selection bias.
\end{remark}

\section{DiD estimators that can highlight or summarize heterogeneity}\label{sec:np}

In this section, we discuss how one can bypass the limitations of the TWFE regression specification in \Cref{eqn:twfe} by proposing data-driven estimation procedures that target well-defined causal parameters. For simplicity, in this section, we rely on \Cref{ass:strong-parallel-trends} so we can get all causal parameters under the same identification assumptions. If one is interested in $\LATT(d|d)$ or their functionals, one can rely on \Cref{ass:continuous-parallel-trends} and use the same estimation procedure for $\ATT(d)$ that we discuss below. In this case, though, we stress that one should not interpret derivatives of estimates of $\LATT(d|d)$ as estimates of $\ACRT(d|d)$.

\subsection{Estimating average causal functions among the treated}\label{sec:np_estimates}
 We start with the estimation of the dose-specific functions, $\ATT(d)$ and $\ACRT(d)$ under \Cref{ass:strong-parallel-trends}. First, recall that, from Theorem \ref{thm:ate}, we have that, for a positive dose $d$,
 \begin{align*}
     \ATT(d) = \E[\Delta Y | D=d] - \E[\Delta Y | D=0],
 \end{align*}
 as well as $\ACRT(d) = {\partial \E[\Delta Y | D=d]}\big/{\partial d}$ when the treatment is continuous, and $\ACRT(d_j) = \left({\E[\Delta Y | D=d_j] - \E[\Delta Y | D=d_{j-1}]}\right)\big/ \left({d_j - d_{j-1}}\right)$ when the treatment is multi-valued discrete. As $\E[\Delta Y | D=0]$ can be estimated using its sample analog, $\E_n[\Delta Y | D=0] = n_{D=0}^{-1}\sum_{i:D_i=0} \Delta Y_i$, with $n_{D=0} = \sum_{i=1}^n 1\{D_i = 0\}$, the main challenge in estimating all these functions resides in estimating $\E[\Delta Y | D=d]$ among treated units ($d>0)$ and its derivative. 
 
 Note that this is a standard regression setup, and, as such, researchers have different options for how to approach it. Examples include adopting a parametric model for $\E[\Delta Y | D=d]$ (e.g., assuming a quadratic model in dose among the treated), or pursuing nonparametric estimators using kernels or sieves/series. We discuss these considerations below. 

For simplicity, we start with setups where the treatment is multi-valued discrete, and takes on a relatively few values. In this case, one can estimate $\ATT(d_j)$ and $\ACRT(d_j)$ for any positive treatment dose $d_j$ in the dose support using a simple saturated regression\footnote{\label{fn:alternative-twfe-discrete-treatment}One can also use the more flexible TWFE regression specification $Y_{i,t} = \sum_{j=1}^J Post_t \cdot \indicator{D_i=d_j}\beta_j + \eta_i + \theta_t + v_{i,t}, \ t=1,2$. We also note that we implicitly take $d_0 = 0$.}
\begin{eqnarray}
            \Delta Y_{i} = \beta_0 + \sum_{j=1}^{J} 1\{D_i=d_j\}\beta_j + \varepsilon_i \label{eqn:multivar_betas},
\end{eqnarray}
where we use the zero treatment dosage as the omitted category. It will then follow that $\widehat{\beta}_j$ and $\left(\widehat{\beta}_j - \widehat{\beta}_{j-1}\right)\big/\left(d_j - d_{j-1}\right)$ are consistent estimators for the $\ATT(d_j)$ and $\ACRT(d_j)$, respectively, and inference procedures are standard. Note that, in this setup, all that our regression \eqref{eqn:multivar_betas} is doing is to automate the appropriate comparison of means justified under our identification assumptions. 

When the dose (among treated units) is continuous, \eqref{eqn:multivar_betas} becomes infeasible. One straightforward estimation approach is to impose a parametric functional form restriction on how $\Delta Y$ varies with $D$ among treated. For instance, one can consider a model in which $\Delta \widetilde{Y}_i =\Delta Y_i - \E_n[\Delta Y | D=0]$ is quadratic in $D$ among treated units, and run the following regression for observations with $D_i>0$\footnote{\label{fn:one-shot-reg}One could also consider the regression $\Delta Y_{i} = \alpha + \indicator{D>0}\big(\beta_0 + \beta_1 D_i+ \beta_2 D_i^2\big) + \varepsilon_i$ for all observations.} %In this specification, one would need to contrast estimated $\beta$'s to get estimates of the $ATT(d)$. }
\vspace{-.2cm}\begin{eqnarray}
            \Delta \widetilde{Y}_i= \beta_0 + \beta_1 D_i+ \beta_2 D_i^2 +\varepsilon_i \label{eqn:parametric_quadratic}.
\end{eqnarray}
When this regression specification is correctly specified, $\widehat{\ATT}_{\text{par}}(d) = \widehat{\beta}_0 + \widehat{\beta}_1 d+ \widehat{\beta}_2 d^2$ and  $\widehat{\ACRT}_{\text{par}}(d) = \widehat{\beta}_1+ 2\widehat{\beta}_2 d$ are consistent estimators for $\ATT(d)$ and $\ACRT(d)$. Pointwise and uniform-in-$d$ inference procedures are standard. Of course, other parametric functional forms can also be adopted.

A limitation of parametric models, such as \eqref{eqn:parametric_quadratic}, is their reliance on potentially incorrect functional form restrictions. In fact, Theorem \ref{thm:twfe-pt} exactly highlights the consequences of misspecification in the linear case. Provided that the sample size is large, however, researchers can use nonparametric procedures to avoid functional form restrictions. This entails considering a nonparametric regression model of $\Delta \widetilde{Y}_i$ on $D_i$ among treated units,
\vspace{-.2cm}\begin{eqnarray}
            \Delta \widetilde{Y}_i= ATT(D_i) +\varepsilon_i \label{eqn:nonparametric}.
\end{eqnarray}
One can estimate \eqref{eqn:nonparametric} in any number of ways. In our application, we have adopted the data-driven nonparametric estimators proposed by \citet{chen-christensen-kankanala-2024}. An appealing feature of this procedure is that it resembles \eqref{eqn:parametric_quadratic} in the sense that, upon computing the optimal sieve-dimension $\widehat{K}$, one runs a linear regression of $ \Delta \widetilde{Y}$ on flexible $\widehat{K}$-dimensional transformations of $D$ (cubic B-splines), $\psi^{\widehat{K}}(D)$, in the subsample of units with $D_i>0$,
\begin{align}
    \Delta \widetilde{Y}_i = \psi^{\widehat{K}}(D)'\beta_{\widehat{K}} + \varepsilon_i,
\end{align}
and then forming the nonparametric estimators for $\ATT(d)$ and $\ACRT(d)$ as
\begin{eqnarray}
    \widehat{\ATT}_{\text{cck}}(d) = \left( \psi^{\widehat{K}}(d)\right)' \widehat{\beta}_{\widehat{K}}, \quad \quad \widehat{\ACRT}_{\text{cck}}(d) =  \left(\partial \psi^{\widehat{K}}(d)\right)' \widehat{\beta}_{\widehat{K}}, \label{eqn:data-driven_estimators}
\end{eqnarray}
where $\partial \psi^{K}(s) = \left(\left.d\psi_{K1}(s)\right/{ds}, \dots, \left.d\psi_{KK}(s)\right/{ds} \right)'$, and $\widehat{\beta}_{\widehat{K}}$ is the ${\widehat{K}}$-dimension vector of OLS estimators for ${\beta}_{\widehat{K}}$.\footnote{As these nonparametric procedures have slower-than-$\sqrt{n}$ rates of convergence, there is no estimation effect from estimating $\E[\Delta Y | D=0]$. } \citet{chen-christensen-kankanala-2024}'s results imply that the nonparametric estimators for  $ATT(d)$ and $ACRT(d)$ curves converge at the fastest possible (i.e., minimax) rate in sup-norm, and lead to uniform confidence bands that are asymptotically narrower (more precise) than those based on undersmoothing, and contract at, or within a $\log \log n$ factor of, the minimax rate. See \Cref{app:cck} for more details on how to compute $\widehat{K}$, as well as how to construct uniform confidence bands based on $\widehat{\ATT}_{\text{cck}}(d)$ and $\widehat{\ACRT}_{\text{cck}}(d)$. Of course, one can adopt other nonparametric estimation and inference procedures and select tuning parameters using alternative criteria, although these may lead to different statistical guarantees. %See, e.g., \citet{Chetverikov2024tuning} for a discussion on selecting tuning parameters. 

\subsection{Estimating summary measures of treatment effects}

Researchers frequently want to report summary estimates to enhance interpretability and/or statistical precision, or because a lower-dimensional parameter is an input into some model or post-estimation calculation. As we showed in Section \ref{sec:baseline}, however, a linear TWFE regression generally fails to deliver an interpretable summary parameter. In this section, we discuss estimation of $\ATT^{\glob}$ and $\ACRT^\glob$, which are summary causal effect parameters that have a clear interpretation.

When there are untreated units, part (a) of \Cref{cor:spt-agg} suggests an extremely simple and familiar estimator of $\ATT^{\glob}$: the difference between the average change in outcomes among treated units minus the average outcome change for untreated units. This ``binarized'' DiD estimator can be obtained from the following simple linear regression specification:
\begin{eqnarray}
            \Delta Y_{i} = \beta_0^{bin} + D_i^{>0}\beta^{bin} + \epsilon_i, \label{eqn:bin_reg}
\end{eqnarray}
where $D_i^{>0} = 1\{D_i > 0\}$. It is straightforward to show that under the identification assumptions in \Cref{cor:spt-agg}, $\beta^{bin} = \ATT^{\glob}$. Note that this estimator applies equally to continuous and multi-valued discrete treatments.

Aggregated average causal response parameters can be constructed easily by weighting the estimated average causal functions across doses using the dose distribution itself. For discrete treatments, it is straightforward to aggregate these $\ACRT(d)$'s based on the coefficients from \eqref{eqn:multivar_betas} to form a plug-in estimator for the $\ACRT^{\glob}$, using the identification formula in \Cref{cor:spt-agg}(c), i.e., 
\vspace{-.1cm}\begin{align}
            \widehat{\ACRT}^{\glob} = \sum_{j=1}^J \dfrac{ \widehat{\beta}_j - \widehat{\beta}_{j-1}}{d_j - d_{j-1}} \widehat{\P}(D=d_j|D>0), \label{eqn:acr_mv}
\end{align}
where $\widehat{\P}(D=d_j|D>0) = \left. \sum_{i=1}^n 1\{D_i = d_j\} \right/\sum_{i=1}^n 1\{D_i > 0\}$. Inference procedures follow from the Delta Method. One can follow a similar strategy when using the scaled $\ATT(d)$ as the ``building block'' of the aggregation. A similar approach applies to estimating $\ACRT^{\glob}$ with a continuous dose. Our proposed estimator is simple to compute as it is based on the plug-in principle, i.e., 
\begin{align*}
\widehat{\ACRT}^{\glob} = \mathbb{E}_n\left[ \left. \widehat{\ACRT}(D)\right| D>0 \right] = \frac{1}{n_{D>0}} \sum_{i: D_i>0} \widehat{\ACRT}(D_i),
 \end{align*}
with $n_{D>0} = \sum_{i=1}^n 1\{D_i>0\}$ denoting the sample size with a positive dose, and $\widehat{\ACRT}(D)$ being a parametric or nonparametric estimator. Under some regularity conditions, one can show that $\widehat{\ACRT}^{\glob}$ is $\sqrt{n_{D>0}}$ consistent and asymptotically normal; see, e.g., Section 4.1 of \citet{ai-chen-2007}.

We close this section by noticing that it is also possible to consider alternative estimators for $\ACRT^{\glob}$ using a so-called Neyman-Orthogonal moment representation. More precisely, by exploring the efficient influence function for $\ACRT^{\glob}$ implied by Theorem 3.1 of \citet{newey-stoker-1993}, it is straightforward to show that 
\begin{align}
             ACRT^{\text{glob}} = \E\left[ACRT(D)  - \left(\Delta Y  - \E[\Delta Y |D, D>0] \right) \dfrac{f'_{D|D>0}(D)}{f_{D|D>0}(D)}\bigg|D>0\right].
\end{align}
Based on this representation, one can then use flexible nonparametric or machine-learning-based estimators for the nuisance functions and still conduct asymptotically valid inference procedures. This opens the door for leveraging double machine learning procedures to estimate $ACRT^{\text{glob}}$ in DiD contexts. We leave this topic for future research.

\section{Extensions} \label{sec:extensions}
In this section, we briefly summarize several extensions of our main results that are further discussed in the Appendix and Supplementary Appendix.

\subsection{Relaxing Strong Parallel Trends}\label{sec:ext_relax_spt}

Under traditional DiD assumptions, \Cref{ass:continuous-parallel-trends} led to the identification of local $\LATT(d|d)$ parameters that are difficult to compare across dosages.  On the other hand, the strong parallel trends assumption led to  $\ATT(d)$ parameters. These can be seen as extreme cases, and it is possible to trade off the strength of assumptions with the type of parameters that can be identified in different ways.  The number of these intermediate possibilities is large, however.  Here, we sketch what we consider to be three main ideas to relax strong parallel trends.  \Cref{app:relaxing-strong-parallel-trends} of the Supplementary Appendix provides substantially more detail.

First, in many cases, researchers may be willing to assume that they know the direction of the selection bias.  For example, suppose that a researcher is willing to assume that, for all $d$ and any dose groups $l < h$, $\LATT(d|l) \leq \LATT(d|h)$, i.e., that higher dose groups would experience larger treatment effects at any value of the dose. In the Supplementary Appendix, we show that this type of assumption leads to bounds on causal effect parameters without requiring strong parallel trends. For example, it implies that, for all $d$
\vspace{-.1cm}\begin{align*}
    \LACRT(d|d) \leq \frac{\partial \E[\Delta Y | D=d]}{\partial d},
\end{align*}
which provides a bound on $\LACRT(d|d)$.  See \Cref{prop:partial-identification} in the Supplementary Appendix for more details.

A second possibility for relaxing strong parallel trends is to define a sub-region $\mathcal{D}_s \subseteq \mathcal{D}_+$ for which strong parallel trends holds, i.e., to assume that 
\begin{align}
    \E[Y_{t=2}(d) - Y_{t=1}(0) | D \in \mathcal{D}_s] = \E[Y_{t=2}(d) - Y_{t=1}(0)|D=d] \label{eqn:local-spt}
\end{align}
holds for all $d \in \mathcal{D}_s$.  Under this assumption, we show in \Cref{prop:local-ate} in the Supplementary Appendix that, for $h,l \in \mathcal{D}_s$, 
\begin{align*}
    \E[\Delta Y | D=h] - \E[\Delta Y | D=l] = \E[Y_{t=2}(h) - Y_{t=2}(l) | D \in \mathcal{D}_s].
\end{align*}
In other words, comparing the trends in outcomes over time for dose group $h$ to dose group $l$ delivers the average causal effect of dose $h$ relative to dose $l$ among those dose groups in $\mathcal{D}_s$.  Under PT, the same comparison would include selection bias terms.  

While the assumption in \Cref{eqn:local-spt} is weaker than SPT, the tradeoff is that now only comparisons within the set $\mathcal{D}_s$ have a causal interpretation.  In some applications, this assumption could be notably weaker than \Cref{ass:strong-parallel-trends}---in fact, this assumption should, at least arguably, no longer be called ``strong parallel trends'' because it is non-trivially non-nested with \Cref{ass:continuous-parallel-trends}.  For example, suppose that $\mathcal{D}_s$ contains large doses.  The assumption in \eqref{eqn:local-spt} says that we can learn about the trend in outcomes for a higher-dose group at a counterfactual lower dose by looking at the trend in outcomes for that lower-dose group, but only for doses in $\mathcal{D}_s$.  This could be much more plausible than Assumption PT, which assumes that even very high dose groups would have experienced the same trend in \textit{untreated} potential outcomes as the untreated group, even though these units might be very different from each other. This local version of the SPT assumption is appealing in applications where there is substantial variation in the dose and the researcher is willing to assume that there is no selection bias among units that select similar doses, but the researcher is unwilling to assume that there is no selection bias among units that select substantially different doses.

Finally, in some applications, strong parallel trends may be more plausible after conditioning on some observed covariates $X$.  Under a version of strong parallel trends conditional on covariates, one can show that the conditional average treatment effect, $\ATT_x(d)= \E[Y_{t=2}(d) - Y_{t=2}(0) | X=x, D>0]$, is identified.  Since this parameter is not local to dose group $d$, conditional on $X=x$, one can compare $\ATT_x(d)$ across different values of the dose without inducing selection bias terms.  This is an intermediate case, however, in that these are more local parameters than $\ATT(d)$ because they are local to the particular value of the covariates $x$.  See the discussion in \Cref{app:relaxing-strong-parallel-trends} in the Supplementary Appendix for more details.

\subsection{Multiple time periods and variation in treatment timing}\label{sec:ext-staggered}

Although our results so far focus on two-period cases, we can extend them to setups with multiple time periods and variation in treatment timing across units by combining the ideas discussed in \Cref{sec:identification} with those in \citet{callaway-santanna-2021}. We consider this setting in detail in \Cref{sec:multi-period}.  

In a setting with staggered treatment adoption (i.e., where once a unit becomes treated with dose $d$, that unit remains treated with dose $d$ in subsequent periods), knowing the time period that a unit becomes treated with a positive dose (which we denote by $G_i$ and refer to as a unit's \textit{timing group}) and dose $D_i$ (i.e., dose group) fully characterizes a unit's sequence of treatments across all periods.  In this context, we need to augment our potential outcomes terminology and write $Y_{i,t}(g,d)$ as the potential outcome of unit $i$ at time $t$ if it were first treated in period $g$, with dose $d$; we write $Y_{i,t}(0)=Y_{i,t}(\infty, 0)$ to denote a unit's untreated potential outcome---the potential outcome in time period $t$ if that unit did not participate in the treatment in any available period. With this notation at hand, we can define a multi-period analog of $\LATT(d|d)$ as
\vspace{-.2cm}\begin{align*}
    \LATT(g,t,d|g,d) = \E[Y_t(g,d) - Y_t(0) | G=g, D=d]~~\text{and}~~\LACRT(g,t,d|g,d) = \frac{ \partial \LATT(g,t,l|g,d)}{ \partial l} \Bigg|_{l=d}
\end{align*}
which are the average treatment effect and average causal response in period $t$ of (i) becoming treated in period $g$ and (ii) experiencing dose $d$ among those in timing group $g$ and dose group $d$. 

Under no anticipation and a multiple-period version of the parallel trends assumption, we show in \Cref{sec:multi-period} that, in post-treatment periods (i.e., periods where $t \geq g$)
\vspace{-.2cm}\begin{align} \label{eqn:latt-mp}
    \LATT(g,t,d|g,d) = \E[Y_t - Y_{g-1} | G=g, D=d] - \E[Y_t - Y_{g-1} | G=\infty, D=0].
\end{align}
The argument is similar to the two-period case discussed earlier.  The main difference is that the expression above involves the ``long difference'' in changes in outcomes over time, i.e., from period $g-1$ to $t$.  The reason for this difference is that $g-1$ is the most recent period for which units in group $g$ were untreated. The expression above uses the never-treated group ($G=\infty$) as the comparison group, but, like the case with a binary treatment, one can use alternative comparison groups such as the not-yet-treated.  Under a multiple-period version of the strong parallel trends assumption, one can take the derivative of the right-hand side of \Cref{eqn:latt-mp} with respect to $d$ to identify $\ACRT(g,t,d|g,d)$. 
 
One complication that arises in the staggered case is that $\LATT(g,t,d|g,d)$ and $\LACRT(g,t,d|g,d)$ are often relatively high-dimensional objects that can be hard to report (and perhaps hard to estimate precisely).  In \Cref{sec:multi-period}, we discuss two main strategies for aggregating these parameters into lower-dimensional objects.  First, we average across timing groups and time periods to target causal effect parameters that are a function of only the dose: $\LATT^{dose}(d|d)$, and $\LACRT^{dose}(d|d)$---these parameters highlight heterogeneous effects across different doses and are analogous to $\LATT(d|d)$ and $\LACRT(d|d)$ in the two-period case that we have emphasized above. They can be averaged across the dose to deliver scalar summary parameters.  Second, we consider event-study parameters:  $\LATT^{es}_{\loc}(e)$, and $\LACRT^{es}_{\loc}(e)$ that average across the dose and highlight how treatment effects and/or causal responses vary with the length of exposure to the treatment---these parameters are the event study analog of $\LATT^{\loc}$ and $\LACRT^{\loc}$ in the two period case above.  See \citet{callaway-goodman-santanna-2024b} for alternative, intermediate aggregations.  The discussion here focuses on causal effect parameters that are local to a specific dose group and timing group, but, like the two-period case discussed above, it is also possible to recover causal effect parameters across all treated units under strong parallel trends; see \Cref{app:mp-more-details} in the Supplementary Appendix for more details.

\subsection{Interpreting TWFE Regressions with Multiple Periods/Groups}
In \Cref{app:twfe-mp} of the Supplementary Appendix, we also extend our TWFE decomposition results from \Cref{thm:twfe-pt} to cover setups beyond the two-period case, including setups with staggered treatment adoptions with continuous or multi-valued discrete treatments. These results generalize the decompositions in \citet{chaisemartin-dhaultfoeuille-2020,goodman-bacon-2021} to the case of a continuous treatment.  Those results demonstrate that TWFE regressions with multiple periods and variation in treatment timing (i) continue to suffer from the weighting and selection bias issues that we highlighted in \Cref{thm:twfe-pt}, (ii) inherit weighting issues (including possible negative weights) that are prevalent in TWFE regressions with binary, staggered treatment adoption, and (iii) are affected by violations of parallel trends in pre-treatment periods.
 
\subsection{Event-Study and Pre-Treatment Differences}\label{sec:ext_pre_treatment}

When there are multiple periods of data available, DiD applications typically assess the plausibility of the parallel trends assumption by checking whether or not parallel trends holds in pre-treatment periods.  In a setting with a continuous treatment, one can check whether $\E[\Delta Y_t|D=d] = \E[\Delta Y_t|D=0]$ is approximately correct for all pre-treatment time periods $t$ and all $d$; one can also check $\E[Y_t - Y_{g-1}|D=d] = \E[Y_t - Y_{g-1}|D=0]$, with $g$ being the time of treatment adoption (where we simplify and consider a single treatment date setup). Implementing these tests, however, can be complicated because it involves multiple dose-response nonparametric estimates.  A convenient alternative is to report aggregated event study parameters such as $\ATT^{es}_{\loc}(e)$ or $\ACRT^{es}_{\loc}(e)$ in pre-treatment periods (i.e., $e<0$). Plotting estimates of $\ATT^{es}_{\loc}(e)$ and $\ACRT^{es}_{\loc}(e)$ for pre-treatment periods ($e<0$) can be used to assess the plausibility of parallel trends. We report these for our empirical application in Figures \ref{fig:F7} and \ref{fig:F9_acr}. Having said that, we note that one possible drawback of this test is that it may overlook violations of the parallel trends assumption that these event-study versions of the test would not detect.

An interesting (though subtle) caveat is that in cases where an aggregate level effect such as $\LATT^{\loc}$ or its event study version $\LATT^{es}_{\loc}(e)$ is the target parameter of the analysis,  it is possible to recover it under ``weaker'' parallel trends assumptions that allow for violations of parallel trends where the \textit{average} violation of parallel trends across dose groups is equal to zero (rather than the violation of parallel trends being equal to zero for \text{all} dose groups)---we refer to the corresponding averaged version of parallel trends as aggregate parallel trends.  If one maintains aggregate parallel trends, then only $\LATT^{es}_{\loc}(e)$ (and not, e.g., $\ACRT^{es}_{\glob}(e)$) is relevant for assessing its plausibility using pre-treatment periods.  That being said, it is debatable whether or not the violations of parallel trends that can be allowed for under aggregate parallel trends should be counted as evidence against the design.

\section{Continuous DiD in Practice: Causal Effects of Medicare PPS}\label{sec:application}
We have so far shown that the causal question of interest shapes identification in a continuous DiD design and argued that it should guide the estimation approach, too. We now apply our preferred average level treatment effect and average causal response estimators to \citet{acemoglu-finkelstein-2008}'s study of Medicare PPS. To map their setting to our theoretical analysis, we consider the balanced panel data component of \citet{acemoglu-finkelstein-2008}, which comprises 5881 hospitals, and also average all pre-treatment outcomes and post-treatment outcomes over time. Thus, we use $t=1$ to denote the average of pre-treatment periods (1980-1983), and $t=2$ to denote the average of post-treatment periods (1984-1986). We also denote treatment dose here by $M$ instead of $D$, as $M$ is a short-hand notation for the 1983 Medicare inpatient share that determines treatment exposure in the AF application.

To begin, consider the profit maximization problem for a hospital with Medicare inpatient share $M$. We follow AF and assume a production function, $F_t(L,K)$, that is homothetic in labor ($L$) and capital ($K$). Market wages and rental rates are normalized by the output price, and Medicare subsidies mean that net input prices are $(1-s_{L,t}M)w$ and $(1-s_{K,t}M)r$. Firms consider the following profit maximization problem:
\vspace{-.2cm}\begin{align*}
\max_{L,K} F_{t}(L,K)-(1-s_{L,t}M)wL-(1-s_{K,t}M)rK.
\end{align*}
The solution to this problem generates factor demands and a capital-labor ratio that is only a function of the input price ratio, $k^*_{t}\bigg(\frac{(1-s_{L,t}M)w}{(1-s_{K,t}M)r}\bigg)$. We write the subsidy ratio, $\frac{(1-s_{L,t}M)}{(1-s_{K,t}M)}$ as $1+S_t(M) = 1+\frac{(s_{K,t}-s_{L,t})M}{1-s_{K,t}M}$. This reflects the fact that hospitals with no Medicare patients ($M=0$), and all hospitals before PPS (when $s_{K,t=1}=s_{L,t=1}=s$) face no relative price distortion. PPS set $s_{L,t}=0$ in 1983, making $S_{t=2}(M) = \frac{s_{K,t=2}M}{1-s_{K,t=2}M}$. 

This structure allows us to define the capital-labor ratio potential outcomes in terms of Medicare inpatient share $M$:
\vspace{-.4cm}\begin{align*}
Y_{t=1} = Y_{t=1}(0)  = k^*_{t=1}\bigg(\frac{w}{r}\bigg) \quad \text{ and } \quad
Y_{t=2} = Y_{t=2}(M) = k^*_{t=2}\bigg(\big(1+S_{t=2}(M)\big)\frac{w}{r}\bigg).
\end{align*}
Three details of the theoretical setup are worth noting. First, homotheticity allows us to connect potential outcomes as a function of $M$ to a firm's optimal capital-labor ratio as a function of relative prices (as a function of $M$).  Without this assumption, a hospital's scale affects its input mix, and capital-labor ratios are a function of net labor and capital prices separately, complicating the theoretical interpretation of causal parameters. Second, we define our parameters of interest in terms of causal effects of $M$ on $Y$. A structural interpretation of those parameters in terms of $k^*$ necessarily involves the non-linear way in which $M$ changes the subsidy ratio, $S_t(M)$ (as well as a kind of exclusion restriction that rules out direct effects of $M$ on outcomes). Third, we use time subscripts to match the fact that PPS changed over time, but this is not a dynamic model. The assumed lack of forward-looking behavior implies the no anticipation assumption (\Cref{ass:no-anticipation}) and allows us to write $Y_{t=1}=Y_{t=1}(0)$. All these details are in line with AF's theoretical model.

\subsection{Causal Questions Around Medicare PPS}
AF are primarily interested in the question: did PPS raise capital-labor ratios? PPS sought to help hospitals invest in new medical technologies with the aim of improving patient outcomes \citep{ota_1984}. But regulators also worried about the ``incentive for hospitals to adopt expensive capital equipment that reduces operating costs but raises total costs per case'' \citep[14]{ota_1984}. Thus, Medicare's role in technology investments has important policy implications. Moreover, the theoretical model predicts that PPS would raise capital-labor ratios for all treated hospitals, so the sign of its effects is a test of a simple neoclassical production theory. The building block parameters that answer these questions are the average treatment effect of PPS on hospitals with $M=m$:
\begin{eqnarray*}
    \LATT(m|m) = \E[Y_{t=2}(m) - Y_{t=2}(0)| M=m] = \E\bigg[k^*_{t=2}\bigg((1+S_{t=2}(m))\frac{w}{r}\bigg) - k^*_{t=2}\bigg(\frac{w}{r}\bigg) \bigg| M=m\bigg].
\end{eqnarray*}
Estimating and plotting the entire $\LATT(m|m)$ function shows which hospitals responded most to PPS and tests the prediction that \textit{all} treated hospitals increase their capital intensity. Under parallel trends alone, one cannot compare across $\LATT(m|m)$'s, as it is not possible to discern whether variation from $\LATT(m|m)$ comes directly from subsidy differences or from treatment effect heterogeneity. Averaging this function across treated hospitals yields $\LATT^{\loc} = \E[\LATT(M|M)|M>0]$, a summary parameter that directly answers the question ``did PPS raise capital-labor ratios on average?''

One may also be interested in which subsidy levels have larger causal effects. For example, if technologies are ``lumpy'', then hospitals may not respond to subsidies too small to cover the minimum investment costs. Improving the design of input subsidies thus requires causal estimates of the responsiveness to different subsidy levels. The causal effects of marginal changes in the subsidy ratio also represent another test of the theoretical model because they are proportional to a hospital's elasticity of substitution, $\sigma_{i,t}(m) = \frac{k^{*\prime}_{i,t}}{k^{*}_{i,t}}\times(1+S_t(m))\times\frac{w}{r}$, which, with two inputs, must be positive. The building block parameters that answer these questions are the average causal responses of PPS:% Note that we motivate these questions with $\ACRT(m)$ parameters because our theoretical results showed that $\LACRT(m|m)$ parameters are not identified under PT and are interpretable as population parameters under SPT.} %Furthermore, we impose regularity conditions that allow one to interchange the order of integration and differentiation.}
\begin{small}
\begin{align}\label{eqn:acr_elast_subs}
  \ACRT(m) = \E\big[ Y^\prime_{t=2}\big(m\big) \big| M>0 \big] &= \E\bigg[ k^{*\prime}_{t=2}\bigg((1+S_{t=2}(m))\frac{w}{r}\bigg)S^\prime_{t=2}(m)\frac{w}{r}\bigg| M>0\bigg] \nonumber \\ 
  &=\E\bigg[\sigma_{{t=2}}(m)k^{*}_{t=2}\bigg((1+S_{t=2}(m))\frac{w}{r}\bigg)\frac{s_k}{1-s_km}\bigg| M>0 \bigg]
\end{align}
\end{small}
Again, reporting estimates of the entire $\ACRT(m)$ function highlights heterogeneity in how hospitals respond to subsidies, and the summary parameter $\ACRT^{\glob}$ provides a single measure of how much hospitals respond on average to small subsidy differences. 

\begin{figure}[ht]
   \begin{center}
    
    \caption{{Changes in Capital-Labor Ratios before and after 1983 versus the Medicare Inpatient Share }}
    \includegraphics[width=.75\textwidth, keepaspectratio]{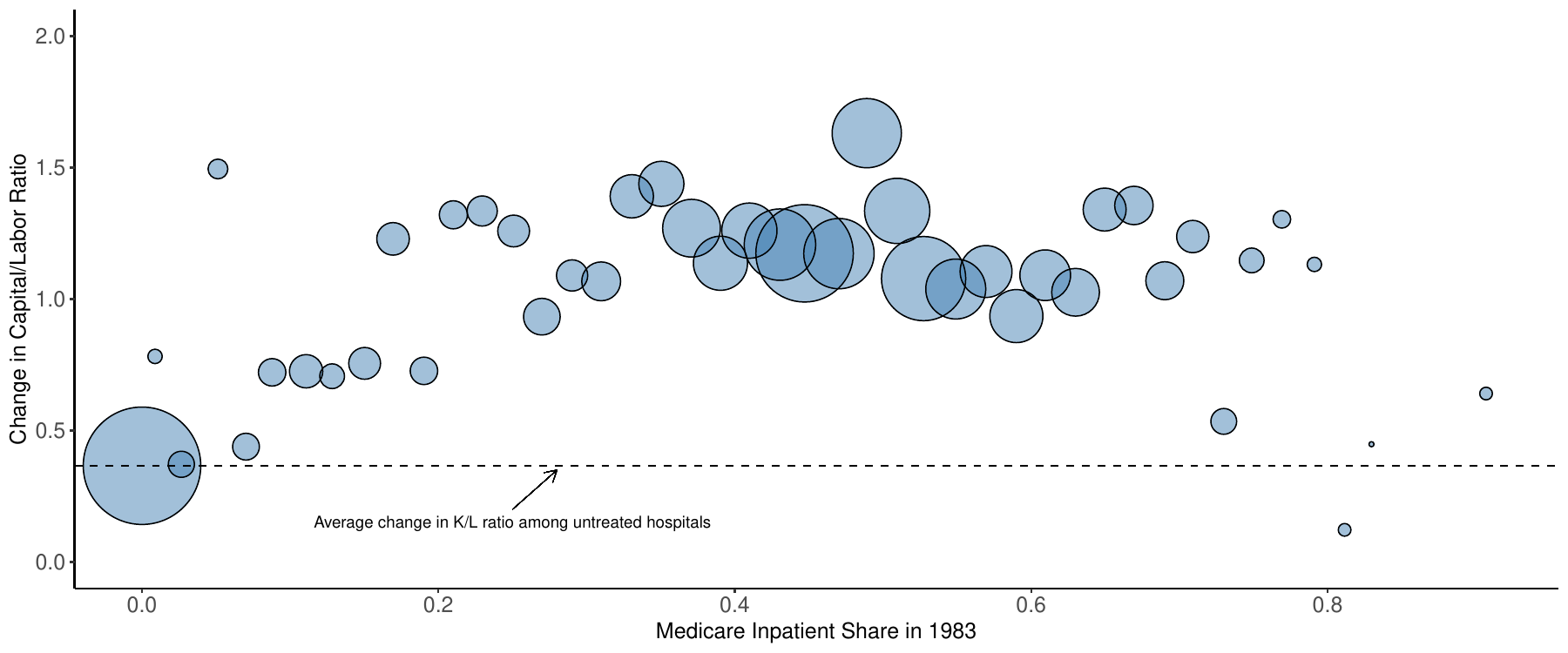}
   \label{fig:F4}
   \end{center}
   \justifying
   \setstretch{.75}
    {\vspace{-.5cm}\scriptsize  \textit{Notes:} The figure presents a binned scatter plot of the change in the average depreciation share (capital-labor ratio) between the periods 1980-1983 and 1984-1986 for hospitals in 2-percentage-point bins of the 1983 Medicare inpatient share, $M$. In the lowest bin, hospitals with $M=0$ are plotted separately from hospitals with $M\in(0,0.02]$. We also consider a single bin for all hospitals with $M>0.84$.}
    %\end{flushleft}
\end{figure}

Before turning to our formal estimates, \Cref{fig:F4} presents a binned scatter plot of the change in mean capital-labor ratios before (1980-1983) and after (1984-1986) PPS against the Medicare share of inpatient days in 1983, $m$. Following AF, we measure the capital-labor ratio using the depreciation share of total costs.

The horizontal line equals the mean change in capital-labor ratio for untreated hospitals (0.37). Each circle is the mean outcome change for a given bin of the Medicare inpatient share, with its size proportional to the number of hospitals in that bin. Almost all groups of treated hospitals had stronger growth in capital intensity than untreated hospitals, consistent with the theoretical prediction. The relationship %does not appear to be monotonic, 
is nonlinear, however, which indicates heterogeneity in average treatment effects, at least, and perhaps heterogeneity in the sign of average causal responses. 

\subsection{Average Treatment Effects of PPS}
\begin{figure}[!ht]
   \begin{center}
    %\captionsetup{margin=10pt,labelfont=bf,position=top}
      \caption{{Nonparametric Estimates of $ATT(m|m)$ for Medicare PPS}}
    \includegraphics[width=.67\textwidth, keepaspectratio]{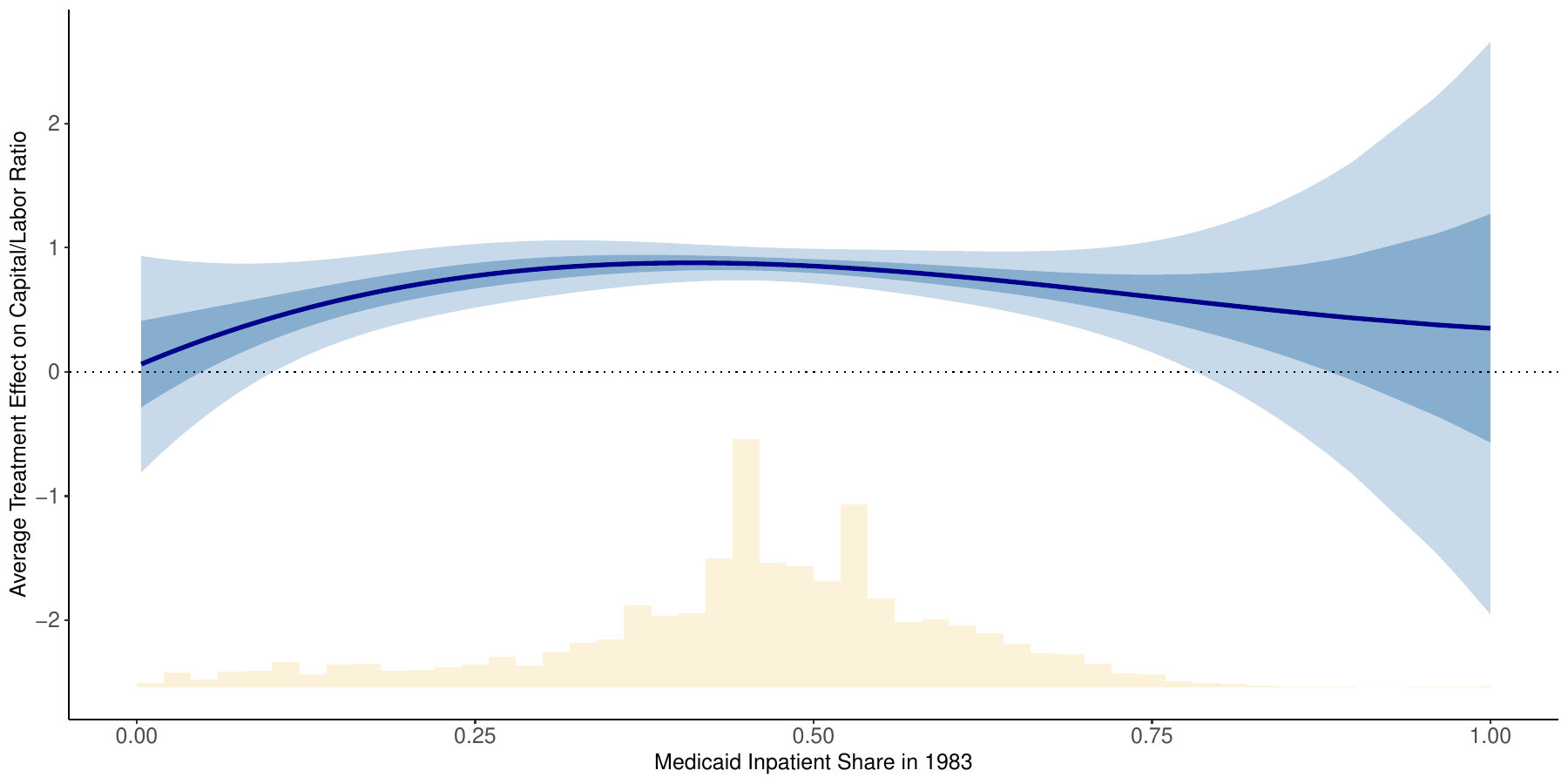}
   \label{fig:F5}
   \end{center}
   \justifying
   \setstretch{.75}
    {\vspace{-.5cm}\scriptsize  \textit{Notes:} The figure plots nonparametric estimate of $\LATT(m|m)$ that adapts the \citet{chen-christensen-kankanala-2024} data-driven estimator to our context, as discussed in Section \ref{sec:np_estimates} and \Cref{app:cck}. The dark-shaded region is the 95-percent point-wise confidence interval, and the lighter-shaded region is the 95-percent honest and sup-norm rate-adaptive uniform confidence band. We display the histogram of the treatment dose among the treated in yellow.}
    %\end{flushleft}
\end{figure}

\Cref{fig:F5} presents our proposed data-adaptive nonparametric estimates of $\LATT(m|m)$ based on \eqref{eqn:data-driven_estimators}.  For inference, we cluster at the hospital level. Our data-driven procedure to optimally choose the sieve dimension selected $\widehat{K}=4$.  These estimates formalize what the scatter plot suggests: that $\LATT(m|m)$ is positive. We plot pointwise 95\% confidence intervals in the dark-shaded region and the wider (honest) 95\% uniform confidence bands in the light-shaded region. We do not detect an effect for values of $m$ below 5 percent, but we reject zero for doses between 0.05 and 0.78, which contains 96 percent of treated hospitals. Significant values of $\widehat{\LATT}(m|m)$ range from about 0.44 percentage points at $m=0.1$ to 0.88 percentage points at $m=0.41$. The average across all doses ($\widehat{\LATT}^{\loc}$) is 0.80 (s.e. = 0.05), or about 18 percent of the 1983 mean outcome (measured by the depreciation share) of 4.5. This evidence suggests that PPS substantially raised capital-labor ratios.

\begin{figure}[ht!]
   \begin{center}
    %\captionsetup{margin=10pt,labelfont=bf,position=top}
      \caption{{Parametric Estimates of $ATT(m|m)$ for Medicare PPS using quadratic specification}}
    \includegraphics[width=.67\textwidth, keepaspectratio]{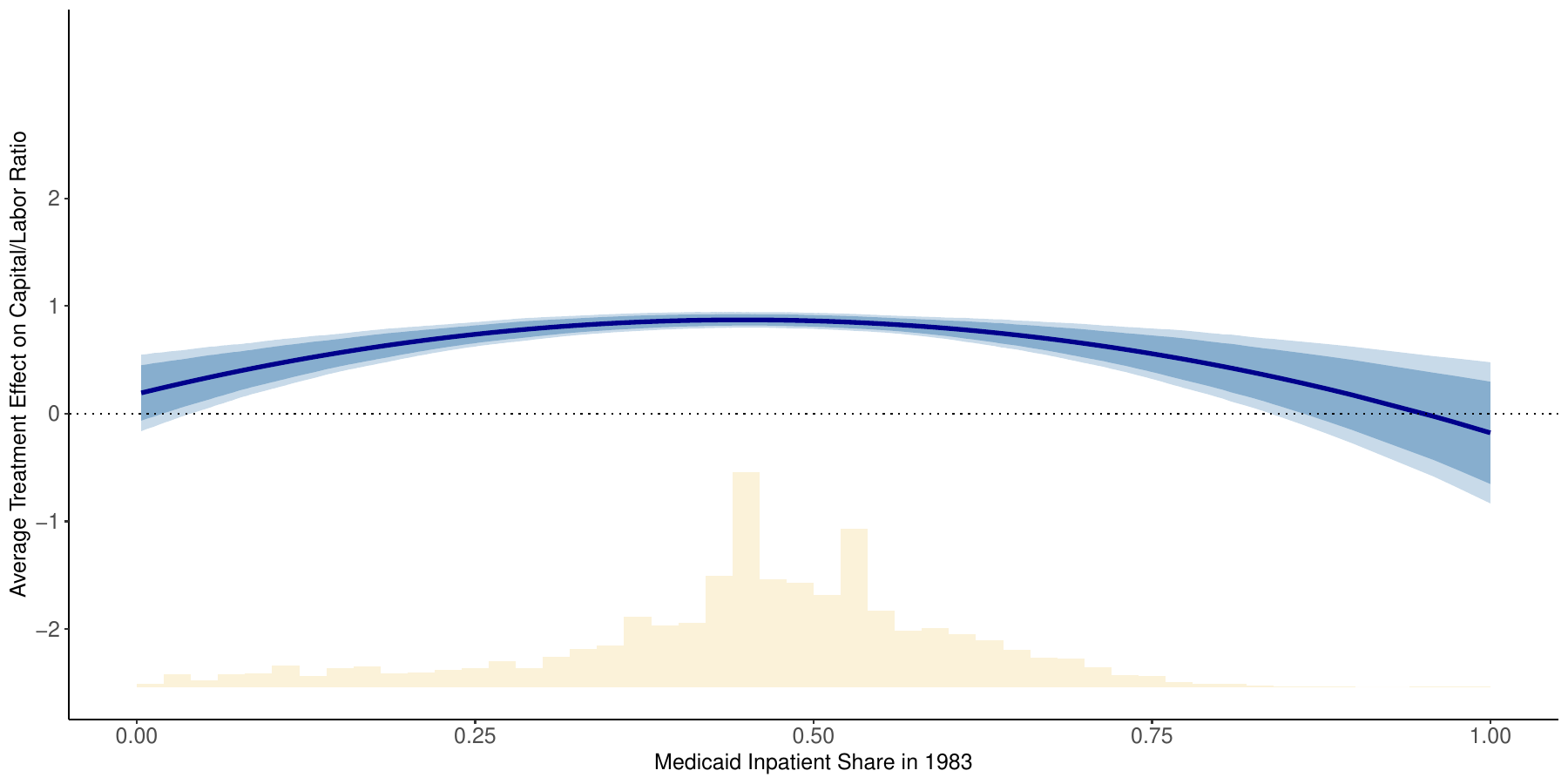}
   \label{fig:F5_parametric}
   \end{center}
   \justifying
   \setstretch{.75}
    {\vspace{-.5cm}\scriptsize  \textit{Notes:} The figure plots parametric estimate of $\LATT(m|m)$ that use the quadratic regression specification in Equation \ref{eqn:parametric_quadratic}.  The dark-shaded region is the 95-percent point-wise confidence interval, and the lighter-shaded region is the 95-percent uniform-in-treatment-dose confidence band. We display the histogram of the treatment dose among the treated in yellow. We use the same y-scale as in \Cref{fig:F5}.}
    %\end{flushleft}
\end{figure}

For comparison, we report in \Cref{fig:F5_parametric} parametric estimates for $\LATT(m|m)$ that use the quadratic regression specification in Equation \ref{eqn:parametric_quadratic}. Different from \Cref{fig:F5}, the interpretability of $\widehat{\LATT}_{\text{par}}(m|m)$ in \Cref{fig:F5_parametric}  depends on the quadratic specification being correctly specified. When we know that is the case, it is clear from \Cref{fig:F5_parametric} that this results in substantially more precise estimates, as they now fully leverage the functional form. Importantly, these gains in precision are more substantial in the regions where data for particular treatment doses are scarce, e.g., for treatment doses above 0.75. Overall, we have 4987 observations with a positive treatment dose. Among these, only 57 have a treatment dose above 0.75, 20 above 0.80, and 3 above 0.90. The rationale for this is very simple: parametric models are good at extrapolating, whereas nonparametric procedures are more cautious about it. The reliability of the extrapolation, once again, crucially depends on the parametric model for $\LATT(m|m)$ being correctly specified.

Although gains in precision are desirable, we caution against using nonparametric results to pick a parametric specification. This, to some extent, resembles a pre-testing problem, and inference based on the parametric model could be misleading. In fact, the appeal of the uniform confidence bands from \citet{chen-christensen-kankanala-2024} that we report in light-shaded blue in \Cref{fig:F5} is that they account for this type of pre-testing issue and are honest, i.e., they are guaranteed to have asymptotically correct coverage over a large (and generic) class of data-generating processes. The uniform confidence bands in \Cref{fig:F5_parametric} are uniform only in treatment dosage, highlighting that it reflects a narrower type of uncertainty than those in \Cref{fig:F5}. Henceforth, as we find it challenging to ex ante motivate a parametric functional form for $\LATT(m|m)$ using arguments grounded in economic theory, we focus our attention on our nonparametric estimators.

In \Cref{sec:twfe}, we argued that $\beta^{twfe}$ should not be relied upon to summarize level effects.  However, the TWFE coefficient is 1.14---roughly similar to our estimate of $\LATT^{\loc}$.  What accounts for their similarity?  One explanation comes from \Cref{eq:twfe-binarized}.  The numerator compares weighted averages of the paths of outcomes for the ``effective'' treated group (those with above-average doses) to the ``effective'' comparison group (those with below-average doses).  However, in our example, slightly more than half of the weight on paths of outcomes in the effective comparison group falls on hospitals with a positive dose.  This biases $\beta^{twfe}$ downward relative to $\LATT^{\loc}$---our estimate of the numerator in \Cref{eq:twfe-binarized} is 0.60.  In contrast, the ``weighted distance'' between the effective treated and comparison groups in the denominator of \Cref{eq:twfe-binarized} is estimated to be 0.53, and dividing by 0.53 results in $\beta^{twfe}$ being upward biased relative to $\LATT^{\loc}$.  That these two biases work in opposite directions and have similar magnitudes in our particular application happens to result in $\widehat{\beta}^{twfe}$ being fairly close to $\widehat{\LATT}^{\loc}$. Interestingly, though, if we instead were to code a hospital's dose on a scale of 0 to 100, our estimate of $\beta^{twfe}$ shrinks to $0.0114 = 1.14/100$ while our estimate of $\LATT^{\loc}$ remains unchanged.

\begin{figure}[ht!]
   \begin{center}
    %\captionsetup{margin=10pt,labelfont=bf,position=top}
      \caption{{Weighting Schemes for TWFE and Dose Distribution Among Treated}}
    \includegraphics[width=.75\textwidth, keepaspectratio]{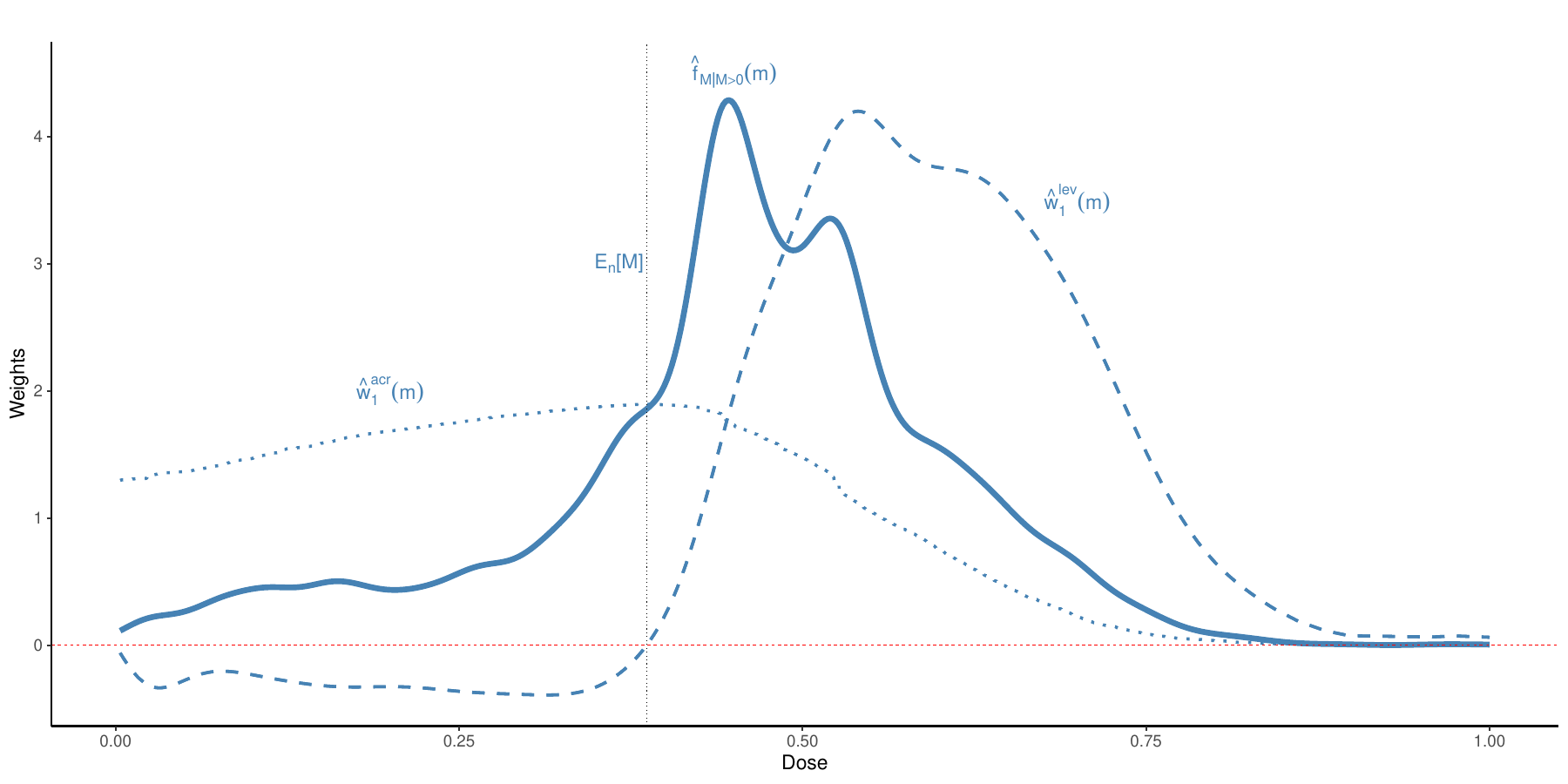}
   \label{fig:F6}
   \end{center}
   \justifying
   \setstretch{.75}
    {\vspace{-.5cm}\scriptsize  \textit{Notes:} The dashed lines are the weights that TWFE puts on $\LATT(m|m)$ and $\ACRT(m)$ parameters, as in \Cref{thm:twfe-pt}. The solid line is a smoothed estimate of the density of the Medicare inpatient share, $M$.}
    %\end{flushleft}
\end{figure}

\begin{figure}[!ht]
   \begin{center}
    %\captionsetup{margin=10pt,labelfont=bf,position=top}
      \caption{{Event-Study Estimates of $ATT$}}
    \includegraphics[width=.75\textwidth, keepaspectratio]{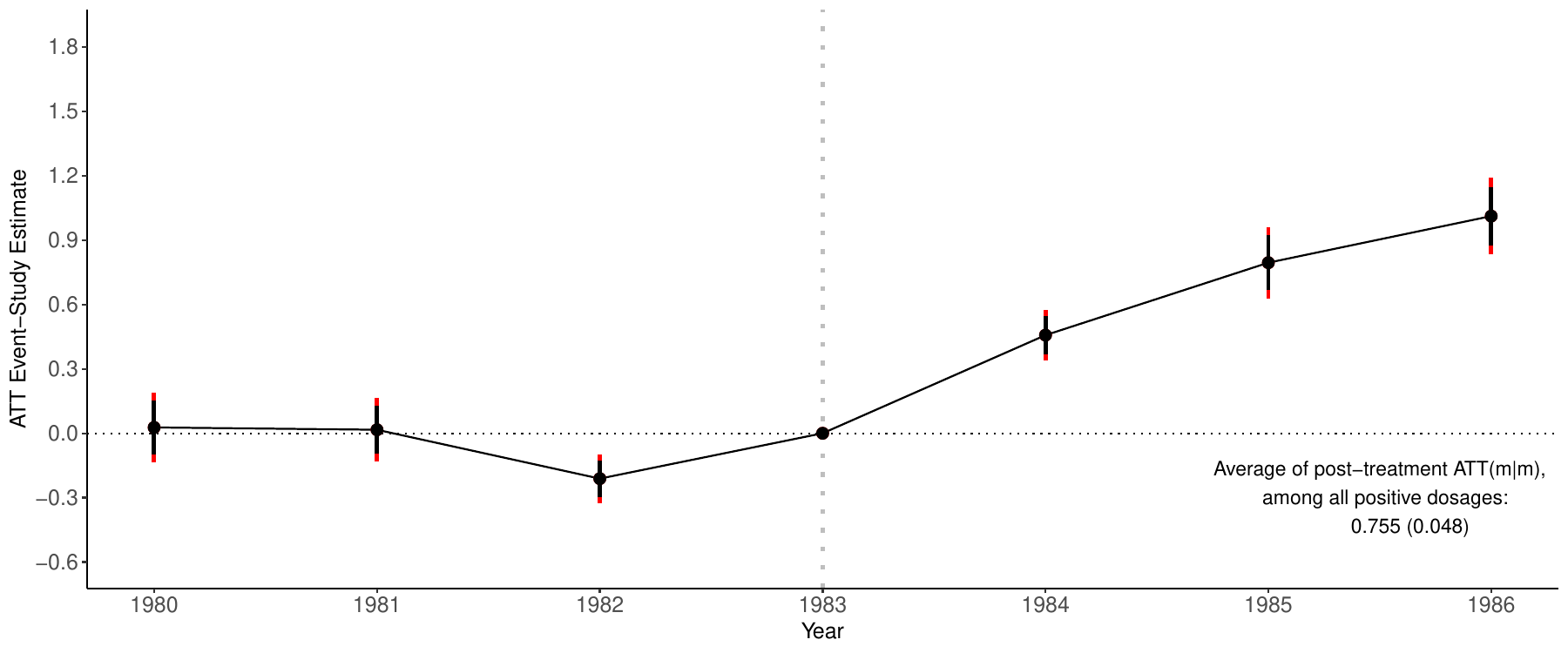}
   \label{fig:F7}
   \end{center}
   \justifying
   \setstretch{.75}
    {\vspace{-.5cm}\scriptsize  \textit{Notes:} The figure plots the event-study estimates of $\LATT^{es}_{\loc}(e)$, with their 95\% pointwise confidence intervals reported in black, and the  95\% uniform confidence bands reported in red.}
    %\end{flushleft}
\end{figure}

Figure \ref{fig:F5} abstracts from dynamics since it is based on average outcomes in the pre- and post-treatment periods. As an alternative, Figure \ref{fig:F7} plots estimates of event-study summary parameters, $\LATT^{es}_{\loc}(e) = \E[Y_{t=e}-Y_{t=1983}|D>0]-\E[Y_{t=e}-Y_{t=1983}|D=0]$, using 1983 as the baseline year.  The patterns are similar to the TWFE event-study in \Cref{fig:F1}, but their magnitudes reflect proper averages of year-specific $\LATT(m|m)$ parameters.\footnote{The negative pre-PPS coefficient may reflect the fact that PPS was passed in April 1983 and partially took effect in that calendar year, and also that hospitals report labor and capital costs for different fiscal years. Therefore, some 1983 outcomes may include post-treatment months. The results also show that the $\LATT^{es}_{\loc}(e)$ grows each year following PPS, which matches the fact that PPS' subsidy reforms actually phased in over three years. We also note, however, that these can represent other types of violations of parallel trends.}

\subsection{Average Causal Responses to PPS}
\begin{figure}[ht]
   \begin{center}
    %\captionsetup{margin=10pt,labelfont=bf,position=top}
      \caption{{Nonparametric Estimates of $ACRT(m)$ for Medicare PPS}}
    \includegraphics[width=.75\textwidth, keepaspectratio]{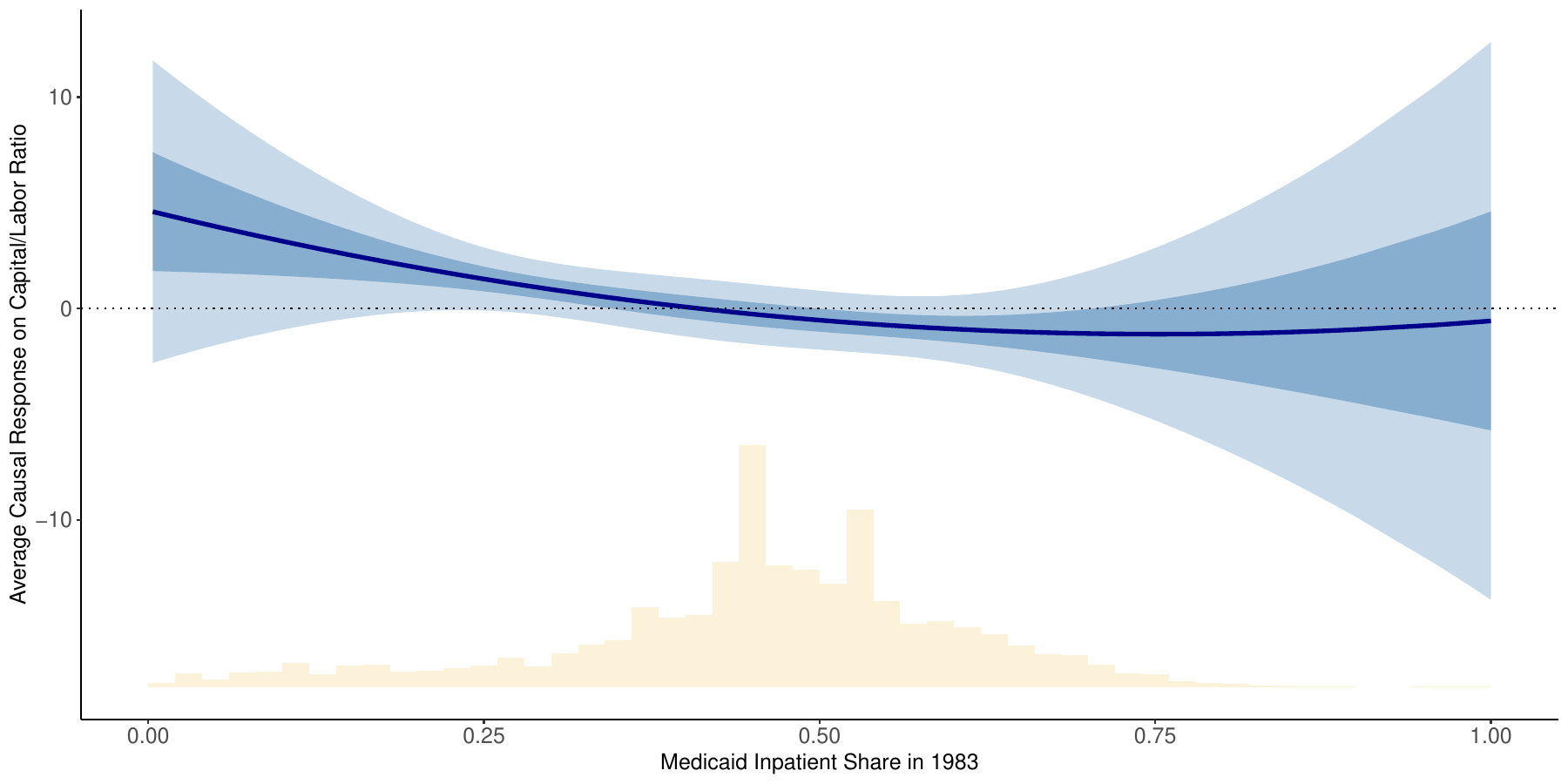}
   \label{fig:F8}
   \end{center}
   \justifying
   \setstretch{.75}
    {\vspace{-.5cm}\scriptsize  \textit{Notes:} The figure plots nonparametric estimate of $\ACRT(m)$ that adapts the \citet{chen-christensen-kankanala-2024} data-driven estimator to our context, as discussed in Section \ref{sec:np_estimates} and \Cref{app:cck}. The dark-shaded region is the 95-percent point-wise confidence interval, and the lighter-shaded region is the 95-percent honest and sup-norm rate-adaptive uniform confidence band. We display the histogram of the treatment dose among the treated in yellow.}
    %\end{flushleft}
\end{figure}

\Cref{fig:F8} plots our proposed data-adaptive nonparametric estimate of the slope of the function estimated in \Cref{fig:F5}. Under \Cref{ass:strong-parallel-trends}, the function in \Cref{fig:F5} is the $\ATT(m)$ and its slope in \Cref{fig:F8} equals the $\ACRT(m)$. The hump shape in \Cref{fig:F5} is reflected in an $\ACRT(m)$ function that starts positive, and declines through most of its support. We estimate negative $\ACRT(m)$ parameters for doses above $m=0.41$, a range that includes 71 percent of treated hospitals. The 95\% uniform confidence band covers zero everywhere, although we are able to detect positive $\ACRT(m)$ values for doses below the mean as well as negative $\ACRT(m)$ values for doses between about 0.5 and 0.7 using pointwise confidence intervals. 

PPS' average causal response parameter weighted by the actual dose distribution of treated hospitals is $\widehat{\ACRT}^{\glob}=-0.08$ (s.e.\,=\,0.19) and is not significantly different from zero.\footnote{We treat the sieve dimension used to compute $\widehat{\ACRT}^{\glob}$ as a non-random sequence, which is in line with the theoretical justification in \citet{ai-chen-2007}. A formal theoretical treatment that accounts for the stochastic nature of our Lepski-type selection is interesting but left for future research.}  This differs substantially from the TWFE coefficient, $\widehat{\beta}^{twfe} = 1.14$.  From \Cref{thm:twfe-pt}(a), the difference between these estimates is fully driven by differences in the weighting scheme.  Our estimate of $\ACRT^{\glob}$ comes from mapping the estimates of $\ACRT(m)$ in \Cref{fig:F8} to the dose distribution weights, $\widehat{f}_{M|M>0}(m)$, in \Cref{fig:F6}; our estimate of $\beta^{twfe}$ comes from mapping the estimates of $\ACRT(m)$ to the TWFE causal response weights, $\widehat{w}_1^{acrt}(m)$, in \Cref{fig:F6}.  As discussed in \Cref{thm:twfe-pt}(a), the TWFE causal response weights are positive for all values of the dose and integrate to one, providing a reason to hope that estimates of $\ACRT^{\glob}$ and $\beta^{twfe}$ would be similar.  However, the TWFE weighting scheme turns out to be much different from the dose distribution weighting scheme.  Combining these differences with the high degree of heterogeneity in $\ACRT(m)$ across $m$ is what leads to the sharp differences in the estimates.  Another reason to emphasize the large difference between these estimates is that the literature has often viewed negative weights as a dividing line between an ``unreasonable'' or ``reasonable'' weighting scheme (see, e.g., \citet{angrist-1998}, \citet{chaisemartin-dhaultfoeuille-2020}, and \citet{blandhol-bonney-mogstad-torgovitsky-2025} for related discussions of this point in different contexts).  The results here suggest that, at least in our context, articulating a well-defined causal effect parameter and targeting that parameter directly is likely to be more important than checking that weights are all positive and integrate to one.

Under Assumption \ref{ass:strong-parallel-trends}, one policy implication of these estimates is that Medicare could have achieved similar, if not greater, capital investments while providing lower capital subsidies. \Cref{fig:F8} shows that marginal increments in the subsidy ratio increase capital intensity only for those with low subsidy levels. Hospitals that received large capital subsidies under PPS responded with smaller increases in capital intensity than hospitals with slightly smaller subsidies, a fact easily seen in the binned scatter plot in \Cref{fig:F4}. The strong parallel trends assumption means that these estimated responses are ``externally valid'' for all treated hospitals, which means that only low subsidies matter for hospitals' input choices. Because higher subsidy ratios do not create further investments in capital, capping capital subsidies may not affect input choices very much.

An important economic implication, however, is that negative $\ACRT(m)$ estimates contradict AF's two-factor economic model. $\ACRT(m)$ is proportional to the average derivative of the optimal capital-labor ratio for hospitals with Medicare share equal to $m$, and Equation \eqref{eqn:acr_elast_subs} shows specifically how it relates to the elasticity of substitution, $\sigma_{i,t}(m)$. To approximate $\E[\sigma_{i,t=2}(m)|M>0]$, we separate out the two terms in \eqref{eqn:acr_elast_subs} and construct $\frac{\ACRT(m)}{\E[Y_{i,{t=2}}|M=m]}\frac{1-s_km}{s_k}$ assuming that $s_k=0.75$. With only two inputs, a rise in the relative price of one must lead to a reduction in its relative use: the elasticity of substitution must be positive. The point estimates of $\E[\sigma_{i,t=2}(m)|M>0]$ do not fit that prediction, although our uniform confidence bands do not reject an average elasticity of substitution of zero. Alternative models, such as a three-factor production function (which AF consider in their working paper) or non-homothetic production, could potentially rationalize this finding.

Finally, both the policy and structural interpretations of \Cref{fig:F8} depend on the strong parallel trends assumption. Without SPT, the slope of $\LATT(m|m)$ may be negative for higher-Medicare-share hospitals simply because their treatment effect functions are systematically lower. Medicare might not have been able to achieve similar capital increases with lower subsidy rates if high-subsidy hospitals just responded differently to low subsidy levels than low-subsidy hospitals did. A negative slope also does not necessarily reject a two-factor production model; just a constant-coefficient model with homogeneous firms, as considered by AF. 

\begin{figure}[!ht]
   \begin{center}
    %\captionsetup{margin=10pt,labelfont=bf,position=top}
      \caption{{Event-Study Estimates of $ACRT$}}
    \includegraphics[width=.9\textwidth, keepaspectratio]{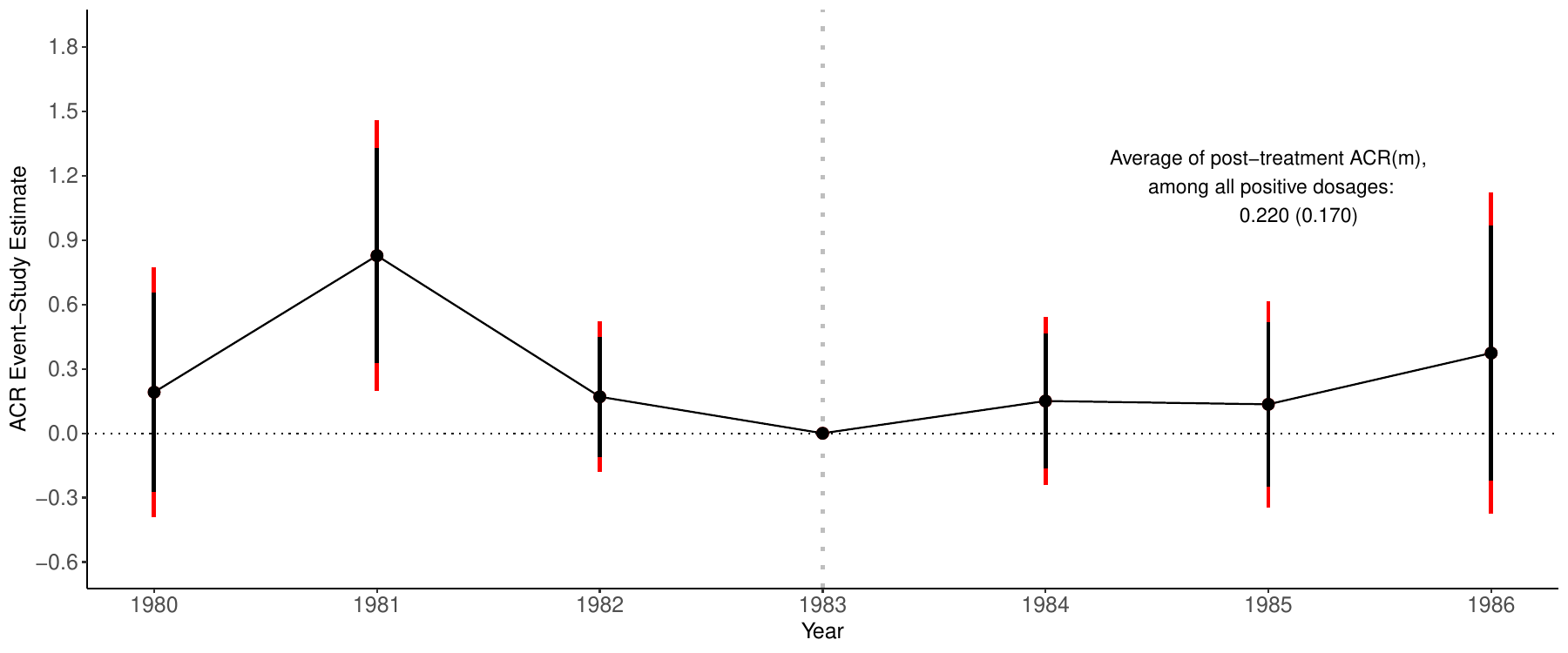}
   \label{fig:F9_acr}
   \end{center}
   \justifying
    {\vspace{-.5cm}\scriptsize  \textit{Notes:} The figure plots the event-study estimates of $\ACRT^{es}_{\glob}(e)$, with their 95\% pointwise confidence intervals reported in black, and the  95\% uniform confidence bands reported in red}
    %\end{flushleft}
\end{figure}

Another indirect way to assess the plausibility of SPT that justifies a causal interpretation of $\ACRT^{\glob}$ is to compute $\ACRT^{es}_{\glob}(e)$, the event-study version of $\ACRT^{\glob}$. These parameters can be estimated using the same procedure discussed in Section \ref{sec:np}, and we plot these in  \Cref{fig:F9_acr}. The no-anticipation assumption means that prior to treatment, when all observed outcomes are untreated potential outcomes, both Assumptions \ref{ass:continuous-parallel-trends} and \ref{ass:strong-parallel-trends} have the same implication: that the average relationship between outcome changes for adjacent dose groups should be zero. Our estimates of these pre-trends reject this in 1981, which is a pre-treatment period. \Cref{fig:F9_acr} corroborates our conclusions about the implausibility of SPT based on implausibly high implied elasticities of substitution.

In summary, our empirical results align with AF's conclusion that the 1983 Medicare reform led hospitals to favor capital over labor.  We find evidence against parallel trends in pre-treatment periods, though the magnitudes of these violations are small relative to estimated effects in post-treatment periods.  Finally, our negative estimates of $\ACRT(m)$ at high values of $m$ cut against the theoretical predictions of the model discussed above; this provides a piece of evidence that casts doubt on the plausibility of strong parallel trends in this application, indicating that one should be cautious when interpreting $ACRT$ parameters.

{
\begingroup
\setstretch{.95}
\renewcommand*{\bibfont}{\small}
\setlength\bibitemsep{5pt}
\printbibliography
\endgroup
}

\appendix

 \section{Proofs of Main Results} \label{app:proofs}

\subsection{Proofs of Results in Section  \ref{sec:identification}}

This section contains the proofs of the results in \Cref{sec:identification} on identifying causal effect parameters such as $\LATT(d|d)$ and $\ATT(d)$ under parallel trends assumptions and with a continuous treatment or multi-valued discrete treatment.

\subsection*{Proof of \Cref{thm:att}}

\begin{proof} To show the result, notice that
\begin{align}
  \LATT(d|d) &= \E[Y_{t=2}(d) - Y_{t=2}(0) | D=d] \nonumber \\
         &= \E[Y_{t=2}(d) - Y_{t=1}(0) | D=d] - \E[Y_{t=2}(0) - Y_{t=1}(0)|D=d] \nonumber \\
         &= \E[Y_{t=2}(d) - Y_{t=1}(0) | D=d] - \E[Y_{t=2}(0) - Y_{t=1}(0)|D=0] \nonumber \\
         &= \E[\Delta Y | D=d] - \E[\Delta Y|D=0] \label{eqn:attdd}
\end{align}
where the second equality holds by adding and subtracting $\E[Y_{t=1}(0)|D=d]$, the third equality holds by \Cref{ass:continuous-parallel-trends}, and the last equality holds because $Y_{t=2}(d)$ and $Y_{t=1}(0)$ are observed potential outcomes when $D=d$ and $Y_{t=2}(0)$ and $Y_{t=1}(0)$ are observed potential outcomes when $D=0$.  That $\LATT^{\loc}$ is identified holds immediately given its definition and that $\LATT(d|d)$ is identified.  To derive the particular expression for $\LATT^{\loc}$, notice that
\begin{align*}
    \LATT^{\loc} &= \E\Big[ \LATT(D|D) \Big| D > 0\Big] \\
    &= \E\Big[ \Big( \E[\Delta Y | D] - \E[\Delta Y | D=0] \Big) \Big| D>0 \Big] \\
    &= \E[\Delta Y | D > 0] - \E[\Delta Y | D=0]
\end{align*}
where the first equality is the definition of $\LATT^{\loc}$, the second equality holds from \Cref{eqn:attdd}, the first part of the third equality holds by an implication of the law of iterated expectations, and the second part of the third equality holds because $\E[\Delta Y | D=0]$ is non-random.
\end{proof}

\subsection*{Proof of \Cref{thm:acr}}

\begin{proof}
To prove part (a), notice that
\begin{align}
    \E[\Delta Y | D=h] - \E[\Delta Y | D=l] &= \Big(\E[\Delta Y | D=h] - \E[\Delta Y | D=0]\Big)  - \Big( \E[\Delta Y | D=l] - \E[\Delta Y | D=0] \Big) \nonumber \\
    &= \LATT(h|h) - \LATT(l|l) \label{eqn:att-diff}
\end{align}
where the first equality holds by adding and subtracting $\E[\Delta Y | D=0]$, and the second equality holds by \Cref{thm:att}.  Next,
\begin{align}
    \LATT(h|h) - \LATT(l|l) &= \E[Y_{t=2}(h) - Y_{t=2}(0) | D=h] - \E[Y_{t=2}(l) - Y_{t=2}(0) | D=l] \nonumber \\
    &= \E[Y_{t=2}(h) - Y_{t=2}(l) | D=h] \nonumber \\
    &~~~+ \E[Y_{t=2}(l) - Y_{t=2}(0) | D=h] - \E[Y_{t=2}(l) - Y_{t=2}(0) | D=l] \nonumber \\
    &= \E[Y_{t=2}(h) - Y_{t=2}(l) | D=h] + \Big(\LATT(l|h) - \LATT(l|l)\Big) \label{eqn:att-diff-causal-response}
\end{align}
where the first equality holds by the definition of $\LATT(d|d)$, the second equality holds by adding and subtracting $\E[Y_{t=2}(l)|D=h]$, and the third equality holds by the definition of $\LATT(l|h)$ and $\LATT(l|l)$.  Notice that $\E[Y_{t=2}(h) - Y_{t=2}(l) | D=h]$ is a causal response of going from dose $l$ to dose $h$ for dose group $h$.  An alternative expression for this term is
\begin{align}
    \E[Y_{t=2}(h) - Y_{t=2}(l) | D=h] = \LATT(h|h) - \LATT(l|h) \label{eqn:att-causal-response}
\end{align} 
Next, we prove part (b).  Using a similar argument as above, notice that, for $d\in \mathcal{D}_+^c$ and $(d+h) \in \mathcal{D}_+^c$, 
\begin{align*}
    \frac{\E[\Delta Y | D=d] - \E[\Delta Y | D=d+h]}{h} &= \frac{\LATT(d|d) - \LATT(d+h|d+h)}{h} \\
    &= \frac{\LATT(d|d) - \LATT(d+h|d)}{h} + \frac{\LATT(d+h|d) - \LATT(d+h|d+h)}{h} 
\end{align*}
where the first equality holds using the same argument as for \Cref{eqn:att-diff}, and the second equality holds by using the arguments in \Cref{eqn:att-diff-causal-response,eqn:att-causal-response}.  The result holds by taking the limit as $h\rightarrow 0$ and the definition of $\LACRT(d|d)$.

Finally, the second result in part (c) involving a discrete treatment holds by taking $h=d_j$ and $l=d_{j-1}$ in \Cref{eqn:att-diff,eqn:att-diff-causal-response} and by the definition of $\LACRT(d_j|d_j)$.
\end{proof}

\subsection*{Proof of \Cref{thm:ate}}
\begin{proof}
For part (a), notice that
  \begin{align*}
    \ATT(d) &= \E[Y_{t=2}(d) - Y_{t=2}(0)|D>0] \\
           &= \E[Y_{t=2}(d) - Y_{t=1}(0)|D>0] - \E[Y_{t=2}(0) - Y_{t=1}(0)|D>0] \\
           &= \E[Y_{t=2}(d) - Y_{t=1}(0)|D=d] - \E[Y_{t=2}(0) - Y_{t=1}(0)|D=0] \\
           &= \E[\Delta Y|D=d] - \E[\Delta Y|D=0]
  \end{align*}
  where the second equality holds by adding and subtracting $\E[Y_{t=1}(0)|D>0]$, the third equality holds by \Cref{ass:strong-parallel-trends}, and the fourth equality holds because $Y_{t=2}(d)$ and $Y_{t=1}(0)$ are observed outcomes when $D=d$.

  Next, we prove the first part of part (b).  First, notice that
  \begin{align*}
      \ATT(h) - \ATT(l) &= \E[Y_{t=2}(h) - Y_{t=2}(0)|D>0] - \E[Y_{t=2}(l) - Y_{t=2}(0)|D>0] \\
      &= \E[Y_{t=2}(h) - Y_{t=2}(l)|D>0]
  \end{align*}
  where the first equality holds by the definition of $\ATT(d)$, and the second equality holds by cancelling the terms involving $Y_{t=2}(0)$.  For the second part,  notice that, from part (a), we have that
  \begin{align*}
      \ATT(h) - \ATT(l) &= \Big(\E[\Delta Y|D=h] - \E[\Delta Y|D=0]\Big) - \Big(\E[\Delta Y|D=l] - \E[\Delta Y|D=0]\Big) \\
      &= \E[\Delta Y|D=h] - \E[\Delta Y|D=l]
  \end{align*}
Now, for part (c), notice that for $d\in \mathcal{D}_+^c$ and $(d+h) \in \mathcal{D}_+^c$, 
\begin{align*}
    \frac{\ATT(d) - \ATT(d+h)}{h} = \frac{\E[\Delta Y | D=d] - \E[\Delta Y | D=d+h]}{h} 
\end{align*}
which follows from part (b).  The result holds by taking the limit as $h \rightarrow 0$ and from the definition of $\ACRT(d)$.  Finally, the result in part (d) involving a discrete treatment holds from part (b) by taking $h=d_j$ and $l=d_{j-1}$ and by the definition of $\ACRT(d_j)$.
\end{proof}

\subsection*{Proof of \Cref{cor:spt-agg}}

\begin{proof}
    The result holds immediately by averaging the results in \Cref{thm:ate} over the distribution of the dose among dose groups that experienced any positive amount of the treatment.
\end{proof}

\section{Adapting CCK to DiD Contexts}\label{app:cck}

In this Appendix, we provide more details on how to adapt the \citet{chen-christensen-kankanala-2024} (henceforth, CCK) data-driven nonparametric estimation and inference procedures in our DiD context. As discussed in  \Cref{sec:np_estimates}, the CCK estimator for $ATT(d)$ and $ACRT(d)$ are given by
\begin{eqnarray*}
    \widehat{\ATT}_{\text{cck}}(d) = \left( \psi^{\widehat{K}}(d)\right)' \widehat{\beta}_{\widehat{K}}, \quad \quad \widehat{\ACRT}_{\text{cck}}(d) =  \left(\partial \psi^{\widehat{K}}(d)\right)' \widehat{\beta}_{\widehat{K}}, 
\end{eqnarray*}
where $\psi^K(d)$ is a $K$-dimensional vector of cubic B-splines basis functions, $\partial \psi^{K}(s) = \left(\left.d\psi_{K1}(s)\right/{ds}, \dots, \left.d\psi_{KK}(s)\right/{ds} \right)'$, $\widehat{\beta}_{\widehat{K}}$ is the ${\widehat{K}}$-dimension vector of OLS estimators for ${\beta}_{\widehat{K}}$, and  $\widehat{K}$ is the CCK data-driven estimator for the optimal sieve dimension. Henceforth, let $n_{D>0} = \sum_{i=1}^n1\{D_i>0\}$ be the sample size with positive treatment dose.

To discuss the optimal choice of the sieve dimension $K$ derived in CCK, we need to add more notation.  Let $\mathcal{K} = \left\{\left(2^k + 3\right): k\in \mathbb{N}_+\cup 0 \right\}$ be the set of possible sieve dimensions for the cubic B-splines. Let $K^+ = \min\{k\in\mathcal{K}:k>K\}$ be the smallest sieve dimension in $\mathcal{K}$ exceeding $K$, and $v_n  =  \max \left\{1, (0.1 \log n)^4 \right\}$. Let $\{\omega_i\}_{i=1}^n$ be iid standard normal draws independent of the data $\{W_i\}_{i=1}^n = \{ Y_{i,t=2}, Y_{i,t=1}, D_i\}_{i=1}^n$. In addition, let $$\widehat{\varphi}_K(W_i, d) = \left(\psi^{K}(d)\right)' \widehat{\phi}_K(W_i),$$ with
\vspace{-.2cm}\begin{align*}
  \widehat{\phi}_K(W_i)  = \mathbb{E}_n\left[1\{D>0\}\cdot \psi^{K}(D)\psi^{K}(D)'\right]^- 1\{D_{i}>0\} \psi^{K}(D_i) \widehat{u}_{i,K},
\end{align*}
and $  \widehat{u}_{i,K}  = \Delta Y_{i} - \E_n[\Delta Y|D=0]  - \left(\psi^{K}(D_i)\right)'   \widehat{\beta}_K$. Finally, for a given $K$ and $K_2$, let 
\vspace{-.1cm}\begin{align*}
 \widehat{\sigma}_{K, K_2}^{2}(d) = \frac{1}{n} \sum_{i=1}^{n} \left( \widehat{\varphi}_K(W_i, d) - \widehat{\varphi}_{K_2}(W_i, d)  \right)^2
 \end{align*}
be an estimator of the (asymptotic) variance of the contrast $\sqrt n \left( \widehat{\ATT}_{K}(d)  - \widehat{\ATT}_{K_2}(d) \right)$, and consider the bootstrap process
\vspace{-.2cm}\begin{align*}
 \mathbb{Z}^{*}_{n}(d,K,K_2) = \dfrac{1}{\widehat{\sigma}_{K, K_2}(d)} \left(\dfrac{1}{\sqrt n} \mathlarger{\sum}_{i=1}^{n}\left( \widehat{\varphi}_K(W_i, d) - \widehat{\varphi}_{K_2}(W_i, d)  \right)\cdot \omega_i\right).
 \end{align*}

For a given sieve dimension $K \in \mathcal{K}$, let
\begin{eqnarray}
    \widehat{\ATT}_{K}(d) = \left(\psi^{K}(d)\right)' \widehat{\beta}_K, \quad \quad \widehat{\ACRT}_{K}(d) =  \left(\partial \psi^{K}(d)\right)' \widehat{\beta}_K, \label{eqn:ate_hat_splines}
\end{eqnarray}
where $\partial \psi^{K}(s) = \left(\left.d\psi_{K1}(s)\right/{ds}, \dots, \left.d\psi_{KK}(s)\right/{ds} \right)'$,
\begin{align}
    \widehat{\beta}_{K} &= \arg \min_{b_K\in\Theta_K} \mathbb{E}_n\left[\left.\left(\Delta Y - \mathbb{E}_n\left[\Delta Y |D=0\right]  - \psi^K(D)'b_K\right)^2 \right|D>0\right] \nonumber \\
    &= \mathbb{E}_n\left[1\{D>0\}\psi^K(D)\psi^K(D)'\right]^-\mathbb{E}_n\left[1\{D>0\}\psi^K(D) \left(\Delta Y - \mathbb{E}_n\left[\Delta Y |D=0\right]\right)\right] \label{eqn:sieve_beta}, 
  \end{align}
and $A^-$ denote the Moore-Penrose inverse of a generic matrix A, and for a generic variable $B$, $$\E_n[B|D>0] = \dfrac{\sum_{i=1}^n 1\{D_i>0\} B_i}{\sum_{i=1}^n 1\{D_i>0\}}.$$

The next algorithm adapts Procedure 1 of CCK to our DiD context and provides the Lepski-type data-driven selection $\widehat{K}$ of the sieve dimension $K$. 
\begin{singlespace}
\vspace{-.1cm}
\begin{algorithm}[Computation of data-driven choice of sieve-dimension $K$ based on CCK.]\label{algo:sieve_dim}
\phantom{a}
    \begin{enumerate}
    \small{ \singlespacing
       \vspace{-.6cm} \item Compute the data-driven index set of sieve dimensions
            \begin{eqnarray}
                \widehat{\mathcal{K}} = \left\{ K \in \mathcal{K}: 0.1\left(\log \widehat{K}_{max}\right)^2 \leq K \leq \widehat{K}_{max} \right\} \label{eqn:Kcal_hat}
            \end{eqnarray}
            \vspace{-.2cm} where 
            \begin{eqnarray}
               \widehat{K}_{max} = \min \left\{ K \in \mathcal{K}: K \sqrt{\log K} v_n \leq 10 \sqrt{n} < K^+ \sqrt{\log K^+} v_n \right\} \label{eqn:Kmax_hat}
            \end{eqnarray}
        \item Let $\alphahat = \min \left\{ 0.5, \sqrt{{\log \Khat_{max}}\big/{\Khat_{max}}} \right\}$. For each independent draw of $\{\omega_i\}_{i=1}^n$, compute 
        \begin{align}
            \sup_{(d,K,K_2)\in \mathcal{D}_+^c\times \widehat{\mathcal{K}} \times \widehat{\mathcal{K}} : K_2 > K}\left| \mathbb{Z}^{*}_{n}(d,K,K_2) \right|.\label{eqn:sup-t-test}
        \end{align}
             Let $\gamma^*_{1 - \alphahat}$ denote the $(1 - \alphahat)$ quantile of  the sup-t statistic \eqref{eqn:sup-t-test} across a large number of independent draws of $\{\omega_i\}_{i=1}^n$, say, 1,000.
        \item The data-driven choice of the sieve dimension is
            \begin{align}
            {\widehat{K} = \inf \left\{ K \in \widehat{\mathcal{K}} ~:~ \sup_{(d,K_2)\in \mathcal{D}_+^c\times \widehat{\mathcal{K}} : K_2 > K} \dfrac{\sqrt n \left| \widehat{\ATT}_{K}(d)  - \widehat{\ATT}_{K_2}(d) \right| }{\widehat{\sigma}_{K, K_2}(d)} \leq 1.1 \gamma^*_{1 - \alphahat} \right\}}. \label{eqn:Khat}
            \end{align}
            }
    \end{enumerate}
\end{algorithm}    
\end{singlespace}

Next, we show how one can form data-driven uniform confidence bands (UCBs) for both $\ATT(d)$ and $\ACRT(d)$ by adapting Procedure 2 of CCK to our DiD context. Toward this end, let $\widehat{A} = \log \log \widehat{K}$ and set $\widehat{\mathcal{K}}_{-} = \{ K \in \widehat{\mathcal{K}} : K<\widehat{K}\}$. Define the bootstrap processes 
\vspace{-.2cm}\begin{align*}
 \mathbb{Z}^{*}_{n}(d,K) = \dfrac{1}{\widehat{\sigma}_{K}(d)} \dfrac{1}{\sqrt n} \mathlarger{\sum}_{i=1}^{n}\widehat{\varphi}_K(W_i, d)\cdot \omega_i, ~~\text{and}~~~~ \mathbb{Z}^{*,acr}_{n}(d,K) = \dfrac{1}{\widehat{\sigma}_{K}^{acrt}(d)} \dfrac{1}{\sqrt n} \mathlarger{\sum}_{i=1}^{n}\widehat{\varphi}_K^{acrt}(W_i, d)\cdot \omega_i.
 \end{align*}
where $\widehat{\varphi}_K^{acrt}(W_i, d) = \left(\partial \psi^{K}(d)\right)' \widehat{\phi}_K(W_i)$,
\vspace{-.2cm}\begin{align*}
 \widehat{\sigma}_{K}^{2}(d) = \frac{1}{n} \sum_{i=1}^{n} \widehat{\varphi}_K(W_i, d) ^2, \quad \quad \text{and} \quad \quad  \widehat{\sigma}_{K}^{acrt,2}(d) = \frac{1}{n} \sum_{i=1}^{n} \widehat{\varphi}_K^{acrt}(W_i, d)^2.
 \end{align*}

\begin{singlespace}
\vspace{-.3cm}
\begin{algorithm}[Computation of UCBs for $ATT(\cdot)$ and $ACRT(d)$ based on CCK.]\label{algo:sieve_UCB}
\phantom{a}
    \begin{enumerate}
    \small{ \singlespacing
      \vspace{-.4cm}  \item[4.] For each independent draw of $\{\omega_i\}_{i=1}^n$, compute 
        \begin{align}
            \hspace{-.1cm}t^* = \sup_{(d,K)\in \mathcal{D}_+^c\times \widehat{\mathcal{K}}_{-}}\left| \mathbb{Z}^{*}_{n}(d,K) \right|,  \quad and  \quad  t^{*,acr} = \sup_{(d,K)\in \mathcal{D}_+^c\times \widehat{\mathcal{K}}_{-}}\left| \mathbb{Z}^{*,acr}_{n}(d,K) \right|.\label{eqn:sup-t-test_UCB}
        \end{align}
             Let $z_{1 - \alpha}^*$ and $z_{1 - \alpha}^{*,acr} $denote the $(1 - \alpha)$ quantile of  the sup-t statistic $t^*$ and $t^{*,acr}$, respectively, across a large number of independent draws of $\{\omega_i\}_{i=1}^n$, say, 1,000.
        \item[5.] The data-driven $100(1-\alpha)\%$ UCB for $\ATT(d)$ and $\ACRT(d)$, $d\in\mathcal{D}_+^c$, are respectively given by
  \begin{small}
    \begin{align}
\hspace{-.1cm}C_{n}(d) &= \left[   \widehat{\ATT}_{\widehat{K}}(d)  - \left(z^{*}_{1-\alpha} + \widehat{A}~\gamma^*_{1 - \alphahat}\right)  \dfrac{\widehat{\sigma}_{\widehat{K}}(d)}{\sqrt{n}}, ~\widehat{\ATT}_{\widehat{K}}(d)  + \left(z^{*}_{1-\alpha} + \widehat{A}~\gamma^*_{1 - \alphahat}\right)  \dfrac{\widehat{\sigma}_{\widehat{K}}(d)}{\sqrt{n}} \right] \label{eqn:UCB} \\
 \hspace{-.1cm} C_{n}^{acrt}(d) &= \left[   \widehat{\ACRT}_{\widehat{K}}(d)  - \left(z^{*,acr}_{1-\alpha} + \widehat{A}~\gamma^*_{1 - \alphahat}\right)  \dfrac{\widehat{\sigma}_{\widehat{K}}^{acrt}(d)}{\sqrt{n}}, ~\widehat{\ACRT}_{\widehat{K}}(d)  + \left(z^{*,acr}_{1-\alpha} + \widehat{A}~\gamma^*_{1 - \alphahat}\right)  \dfrac{\widehat{\sigma}_{\widehat{K}}^{acrt}(d)}{\sqrt{n}} \right] \label{eqn:UCB_acr}
\end{align}
  \end{small}
  }
    \end{enumerate}
\end{algorithm}
\end{singlespace}

\section{Multiple Periods and Variation in Treatment Timing and Dose}\label{sec:multi-period}

DiD applications often use more than two time periods, wherein treatments, whether binary or not, can turn on at different times for different units. This section extends the results from the main text to allow for multiple time periods ($t = 1,...,\T$) with variation in the time when units become treated.  We refer to the time period when a unit becomes treated as a unit's \textit{timing group}, which we denote by $G_i$, which takes values in the set $\mathcal{G}$.  By convention, we set $G=\infty$ for units that remain untreated across all time periods, and we exclude units that are treated in the first period so that $\mathcal{G} \subseteq \{2,\ldots,\T, \infty\}$; we also set $\bar{\mathcal{G}} = \mathcal{G} \setminus \{ \infty \}$ to be the set of all timing groups that ever participate in the treatment.  Treated units receive dose $D = d \in \mathcal{D}_+$.  As in the two-period case, the dose actually experienced, $D$, also defines a unit's \textit{dose group}. % We focus on the case where treatment is an absorbing state, so that the amount of the treatment remains constant in post-treatment periods.  

We extend the potential outcomes notation from the previous section to allow for variation in treatment timing.  Therefore, potential outcomes $Y_{i,t}(g,d)$ denote the outcome for unit $i$ at time period $t$ when such a unit is first treated in period $g$ with dosage $d$.
Note that treated potential outcomes at time $t$ depend on when a unit first becomes treated---i.e., $Y_{i,t}(g,d)$ may not equal $Y_{i,t}(g',d)$ for $g \neq g'$---which allows for general treatment effect dynamics.  $Y_{i,t}(\infty,0)$ is the outcome that unit $i$ would experience if it did not participate in the treatment in any period. We write $Y_{i,t}(0) = Y_{i,t}(\infty,0)$ and refer to this as a unit's untreated potential outcome. %\footnote{There are several empirically relevant settings that are implicitly ruled out by our notation.  For example, the analysis in this section could be extended to allow for units to be ``treated'' at time $g$ but with $d=0$.  This is relevant in applications where, for example, units may live in a jurisdiction that implements a program at time $g$ for which they are not eligible.  Similarly, we could allow for units to have dose $d$ but remain untreated $g=\infty$.  This would make sense if a program's dose was based on a known formula so that it was possible to observe $d$ even for units not actually selected for treatment.  We leave these sort of extensions for future work.} 
We also define the variable $W_{i,t} = D_i \indicator{t \geq G_i}$. which is the amount of dose that unit $i$ experiences in time period $t$; $W_{i,t} = 0$ for all units that are not yet treated by time period $t$.  

Throughout this section, we make the following assumptions.
\begin{namedassumption}{1-MP}[Random Sampling] \label{ass:random-sampling-mp}
    The observed data consists of $\{Y_{i1}, \ldots, Y_{i\T}, D_i, G_i \}_{i=1}^n$ which is independent and identically distributed.
\end{namedassumption}
\begin{namedassumption}{2-MP}[Support] \label{ass:support-mp} \ 

    \vspace{-.5cm}
    \begin{itemize}
        \item[(a)] The support of $D$, $\mathcal{D}=\{0\} \cup \mathcal{D}_+$ where $\mathcal{D}_+ \subseteq (0,\infty)$.  In addition, $\P(D=0) > 0$ and $dF_{D|G}(d|g) > 0$ for all $(g,d) \in \bar{\mathcal{G}} \times \mathcal{D}_+$.
        \vspace{-.3cm}\item[(b)] $\mathcal{D}_+^c = (d_L,d_U) \subset \mathcal{D}_+$.  In addition, for all $g \in \bar{\mathcal{G}}$ and $t = 2,\ldots,\T$, $\E[\Delta Y_t | G=g, D=d]$ is continuously differentiable in $d$ on $\mathcal{D}_+^c$.
    \end{itemize}
\end{namedassumption}
\begin{namedassumption}{3-MP}[No Anticipation / Staggered Adoption] \label{ass:no-anticipation-mp} \ 

    \vspace{-.5cm}
    \begin{itemize}
    \item[(a)] For all $g \in \mathcal{G}$ and $t=1,\ldots,\T$ with $t < g$ (i.e., in pre-treatment periods), $Y_{i,t}(g,d) = Y_{i,t}(0)$.
    \vspace{-.3cm} \item[(b)]
    $W_{i,1} = 0$ almost surely, and, for $t=2,\ldots, \T$ and $d\in \mathcal{D}_+$,  $W_{i,t-1} = d$ implies that $W_{i,t} = d$.
    \end{itemize}
\end{namedassumption}

We next introduce versions of Assumption \ref{ass:continuous-parallel-trends} and \ref{ass:strong-parallel-trends} that are suitable for the setting with multiple periods and variation in treatment timing. \footnote{Besides differences related to multiple periods and variation in treatment timing, the version of strong parallel trends made here is slightly different from \Cref{ass:strong-parallel-trends} in the main text.  Part of the difference comes from there being no untreated units in group $g \in \bar{\mathcal{G}}$, which is why there is a separate part of the assumption for untreated potential outcomes. The other difference is that both parts of the assumption hold for all dose groups rather than on average (i.e., we condition on dose group $l$ in the first part and on dose group $d$ in the second part for untreated potential outcomes).  The version here is stronger and is made for clarity and because we target $\LACRT(g,t,d|g,d)$ rather than $\ACRT(g,t,d)$ in this part of the paper.  }  

\begin{namedassumption}{PT-MP}[Parallel Trends with Multiple Periods and Variation in Treatment Timing] \  \label{ass:parallel-trends-multiple-periods} For all $g \in \bar{\mathcal{G}}$, $t=2,\ldots,\T$, $d \in \mathcal{D}_+$, $\E[\Delta Y_t(0) | G=g, D=d] = \E[\Delta Y_t(0) | G=\infty, D=0]$.
\end{namedassumption}

\begin{namedassumption}{SPT-MP}[Strong Parallel Trends with Multiple Periods and Variation in Treatment Timing] \  \label{ass:strong-parallel-trends-multiple-periods}
    For all $g \in \bar{\mathcal{G}}$, $t=2,\ldots,\T$, and $l,d \in \mathcal{D}_+$, $\E[Y_t(g,d) - Y_{t-1}(g,d) | G=g, D=l] = \E[Y_t(g,d) - Y_{t-1}(g,d) | G=g, D=d]$ and $\E[\Delta Y_t(0) | G=g, D=d] = \E[\Delta Y_t(0) | G=\infty, D=0]$.
\end{namedassumption}

For each unit, we observe their outcome in  period $t$, $Y_{i,t}$, which is given by
\begin{align*}
    Y_{i,t} = Y_{i,t}(0) \indicator{t < G_i} + Y_{i,t}(G_i,D_i) \indicator{t \geq G_i}.
\end{align*}

\subsection{Identification with a Staggered Continuous Treatment}

The causal parameters of interest are the same as in our baseline case, except that they are separately defined for each timing group and in each post-treatment time period:  %The average treatment effect parameters of dose $d$, for group $g$, in time period $t$ are:
\begin{align*}
    \LATT(g,t,d|g,d) &= \E[Y_t(g,d) - Y_t(0) | G=g, D=d],~~\text{and}~~ \ATT(g,t,d) = \E[Y_t(g,d) - Y_t(0) | G=g, D>0].
\end{align*}

Causal response parameters are similarly defined as the effect of a marginal change in the dose on the outcomes of timing group $g$ in period $t$.  For continuous treatments, these are defined as 
\begin{align*}
    \LACRT(g,t,d|g,d) &= \frac{ \partial \LATT(g,t,l|g,d)}{ \partial l} \Bigg|_{l=d} = \frac{ \partial \E\left[Y_t(g,l) | G=g, D=d\right]}{\partial l} \Bigg|_{l=d}, \\%[8pt]
    \ACRT(g,t,d) &= \frac{\partial \ATT(g,t,d)}{\partial d} = \frac{ \partial \E\left[Y_t(g,d) | G=g\right]}{\partial d}.
\end{align*}
For discrete treatments, these are defined as
\begin{align*}
    \LACRT(g,t,d_j|g,d_j) &=  \E[Y_t(g,d_j) - Y_t(g,d_{j-1}) | D=d_j, G=g] \Big/ (d_j - d_{j-1}),\\
    \LACRT(g,t,d_j) &=  \E[Y_t(g,d_j) - Y_t(g,d_{j-1}) | D>0, G=g] \Big/ (d_j - d_{j-1}),
\end{align*}

For brevity, henceforth we focus on the ``local'' causal effect parameters $\LATT(g,t,d|g,d)$ and $\LACRT(g,t,d|g,d)$, which are analogous to the local causal effect parameters $\LATT(d|d)$ and $\LACRT(d|d)$ in the two-period case that we emphasized in the main text.

\begin{theorem} \label{thm:att-dgt}  Under Assumptions \ref{ass:random-sampling-mp}, \ref{ass:support-mp}(a), \ref{ass:no-anticipation-mp}, and \ref{ass:parallel-trends-multiple-periods}, and for all $g \in \bar{\mathcal{G}}$, $t = 2, \ldots, \T$ such that $t \geq g$, and for all $d \in \mathcal{D}_+$,   
    \begin{align*}
        \ATT(g,t,d|g,d) = \E[Y_t - Y_{g-1} | G=g, D=d] - \E[Y_t - Y_{g-1} | W_t=0].
    \end{align*}
If, in addition, Assumptions \ref{ass:support-mp}(b) and \ref{ass:strong-parallel-trends-multiple-periods} hold, then, for all $d \in \mathcal{D}_+^c$,
    \begin{align*}
        \ACRT(g,t,d|g,d) = \frac{\partial \E[Y_t - Y_{g-1} | G=g, D=d]}{\partial d}.
    \end{align*}
\end{theorem}

The proof of \Cref{thm:att-dgt} is provided in \Cref{app:mp-proofs} in the Supplementary Appendix.  The result is broadly similar to the one in the case with two periods.  The first part says that, under \Cref{ass:parallel-trends-multiple-periods}, $\ATT(g,t,d|g,d)$ can be recovered by a DiD comparison between the path of outcomes from period $g-1$ to period $t$ for units in group $g$ treated with dose $d$ and the path of outcomes among units that have not participated in the treatment yet (the setup in this section also rationalizes using the never-treated group, $G=\infty$, as the comparison group as was mentioned in \Cref{sec:extensions}).  Relative to the case with two time periods, the main difference is that the ``base period'' is $g-1$.  The reason for using the base period $g-1$ is that it is the most recent time period when the researcher observes untreated potential outcomes for units in group $g$.  Thus, the result is very much like the case with two time periods: take the most recent untreated potential outcomes for units in a particular group, impute the path of outcomes that they would have experienced in the absence of participating in the treatment from the group of not-yet-treated units (these steps yield mean untreated potential outcomes that units in group $g$ would have experienced in time period $t$) and compare this to the outcomes that are actually observed for units in group $g$ that experienced dose $d$.  The second part says that, under \Cref{ass:strong-parallel-trends-multiple-periods}, $\ACRT(g,t,d|g,d)$ can be recovered by taking the derivative of the average path of outcomes from period $g-1$ to period $t$ among timing group $g$ that experienced dose $d$.  Similarly to the arguments in the main text, if Assumption \ref{ass:parallel-trends-multiple-periods} held rather than \ref{ass:strong-parallel-trends-multiple-periods}, then the same derivative term would additionally include selection bias terms.

Given the results in \Cref{thm:att-dgt}, it follows that causal summary parameters that are aggregations of these dose-and-timing-group-specific parameters are also identified. Of course, one can consider many different types of aggregation, as discussed in \citet{callaway-santanna-2021} and \citet{callaway-goodman-santanna-2024b}, for example. Here, given that the treatment is continuous, we discuss some aggregations that can remain dose-specific and help highlight heterogeneity in treatment dosages. We provide these estimands as being explicitly about them if useful for estimation and inference, and provide more transparency when comparing across procedures \citep{baker-callaway-cunningham-goodman-santanna-2025}.

We start discussing  natural aggregated parameters, including the average treatment effect of dose $d$ across post-treatment periods for dose group $d$,
\begin{align*}
    \LATT^{dose}(d|d) = \E\Big[ \overline{TE}(d) \Big| D=d, G \leq \T \Big]=\sum_{g \in \bar{\mathcal{G}}} \sum_{t=2}^{\T} \omega^{dose}(g,t,d) \LATT(g,t,d|g,d),
\end{align*}
where $\omega^{dose}(g,t,d) = \frac{\indicator{t \geq g}}{\T - g + 1} \P(G=g | D=d, G \leq \T)$, and
\begin{align*}
    \overline{TE}_i(d) = \frac{1}{\T-G_i+1} \sum_{t=G_i}^{\T} \Big(Y_{i,t}(G_i,d) - Y_{i,t}(0)\Big).
\end{align*}
We can likewise define a causal response parameter
\begin{align*}
    \LACRT^{dose}(d|d) = \frac{\partial \LATT^{dose}(l|d)}{\partial l} \Bigg|_{l=d} = \sum_{g \in \bar{\mathcal{G}}} \sum_{t=2}^{\T} \omega^{dose}(g,t,d) \LACRT(g,t,d|g,d),
\end{align*}
and even  further aggregate these parameters into scalar summary parameters: 
\begin{align*}
    \LATT^{\loc} = \E\Big[ \LATT^{dose}(D|D) \Big| G \leq \T \Big] ~~\textrm{and}~~ \LACRT^{\loc} = \E\Big[ \LACRT^{dose}(D|D) \Big| G \leq \T \Big].
\end{align*}

Empirical researchers are also often interested in analyzing how average treatment effects vary with elapsed treatment timing and consider event-study-type parameters. In our context, with continuous and staggered treatments, one can consider the following dose-specific event study parameters,
\begin{small}
\begin{align*}
    \widetilde{\LATT}^{dose,es}(d|d,e) &= \E\Big[ TE(d|e) \Big| D=d, G+e \in [2,\T], G \leq \T \Big] = \sum_{g \in \bar{\mathcal{G}}} \sum_{t=2}^{\T}  \omega^{dose,es}(g,t,d|e) \LATT(g,t,d|g,d) ,\\
    \widetilde{\LACRT}^{dose,es}(d|d,e) &= \frac{\partial \widetilde{\LATT}^{dose,es}(l|d,e)}{\partial l}\Bigg|_{l=d} = \sum_{g \in \bar{\mathcal{G}}} \sum_{t=2}^{\T} \omega^{dose,es}(g,t,d|e) \LACRT(g,t,d|g,d)
\end{align*}
\end{small}
where $\widetilde{\LATT}^{dose,es}(d|d,e)$ and $\widetilde{\LACRT}^{dose,es}(d|d,e)$ are the average treatment effect of dose $d$ and average causal response of dose $d$ among those in dose group $d$ for those that have been exposed to the treatment for $e$ periods, $TE_i(d|e) = Y_{i,G_i+e}(G_i,d) - Y_{i,G_i+e}(0)$, $\omega^{dose,es}(g,t,d|e) = \indicator{g+e \in [2,\T]} \indicator{g+e=t} \pi_g(e,d)$ and $ \pi_g(e,d) = \P(G=g | D=d, G+e \in [2,\T], G \leq \T)$,

When one also wants to aggregate over treatment dosages to get an easier-to-estimate causal parameter of interest, \begin{align*}
    \LATT^{es}_{\loc}(e) &= \E\Big[ \widetilde{ATT}^{dose,es}(D|D,e) \Big| G+e \in [2,\T], G \leq \T \Big] \\
    \LACRT^{es}_{\loc}(e) &= \E\left[ \widetilde{ACRT}^{dose,es}(D|D,e) \Big| G+e \in [2,\T], G \leq \T \right]
\end{align*}
which provide event study versions of average treatment effects and average causal responses across different lengths of exposure to the treatment.  For values of $e \geq 0$, $\ATT^{es}_{\loc}(e)$ and $\ACRT^{es}_{\loc}(e)$ are related to treatment effect dynamics. It is also interesting to consider cases where $e < 0$, which can be interpreted as a pre-test of the parallel trends assumption. See also \citet{callaway-goodman-santanna-2024b} for a discussion.

\begin{remark}
    We do not provide formal estimation results for the setting with multiple periods and variation in treatment timing. However, we note that, if one bases estimation on the sample analog of the results in \Cref{thm:att-dgt}, then the results in the main text for the case with two periods apply directly to the disaggregated parameters $\LATT(g,t,d|g,d)$ and $\LACRT(g,t,d|g,d)$.  
    %For the aggregated parameters discussed above, at a high level, one can then proceed to combine estimation results for the disaggregated parameters with the estimation results for the related aggregation schemes proposed in \citet{callaway-santanna-2021}.  
    Then, one can estimate any of the aggregated parameters discussed above as the appropriate weighted average of $\LATT(g,t,d|g,d)$ or $\LACRT(g,t,d|g,d)$.  Interestingly, given the results in \Cref{cor:spt-agg} and \citet{callaway-santanna-2021}, when $\ATT^{es}_{\loc}(e)$ is the target parameter, one can binarize the treatment (i.e., classify units as being treated if they experience any positive amount of the treatment) and simply rely on the event-study procedures proposed by  \citet{callaway-santanna-2021}.
\end{remark}

\end{document}